\documentclass[preprint,5p]{elsarticle}

\usepackage{graphicx}
\usepackage{dcolumn}
\usepackage{bm}
\usepackage[colorinlistoftodos]{todonotes}
\usepackage[colorlinks=true, allcolors=black]{hyperref}
\usepackage{array,colortbl,xcolor}
\usepackage{siunitx}
\usepackage{caption}
\usepackage{subcaption}
\usepackage[percent]{overpic}
\usepackage{lineno}
\usepackage{amsmath}
\usepackage{fancyvrb,cprotect}

\usepackage{array, boldline, makecell, booktabs}

\begin{document}

\title{Design, Construction, and Performance of the GEM based Radial Time Projection Chamber for the BONuS12 Experiment with CLAS12}

\author[HU]{I.~Albayrak}
\author[cea]{S.~Aune}
\author[ODU,WM]{C.~Ayerbe Gayoso}
\author[cea]{P.~Baron}
\author[ODU]{S.~B\"ultmann} 
\author[ODU,ipn]{G.~Charles}
\author[HU,Jlab]{M.~E.~Christy}
\author[ODU]{G.~Dodge}
\author[ODU]{N.~Dzbenski}
\author[ipn]{R.~Dupr\'e}
\author[WM]{K.~Griffioen}
\author[ODU]{M.~Hattawy\corref{cor}}\ead{mhattawy@odu.edu} 
\author[ODU]{Y.~C.~Hung} 
\author[UVA]{N.~Kalantarians}
\author[ODU]{S.~Kuhn}
\author[cea]{I.~Mandjavidze}
\author[HU]{A.~Nadeeshani}
\author[ipn]{M.~Ouillon}
\author[ODU]{P.~Pandey}
\author[ODU]{D.~Payette}
\author[ODU]{M.~Pokhrel}
\author[ODU,Jlab]{J.~Poudel}
\author[Jlab]{A.S.~Tadepalli}
\author[cea]{M.~Vandenbroucke}

\address[HU]{Hampton University, Hampton, VA 23669, USA}
\address[cea]{Irfu, CEA, Universit\'e Paris-Saclay, 91191, Gif-sur-Yvette, France}
\address[ODU]{Old Dominion University, Norfolk, VA 23529, USA}
\address[WM]{College of William and Mary, Williamsburg, VA 23187, USA}
\address[ipn]{Universit\'e Paris-Saclay, CNRS, IJCLab/IN2P3, 91405, Orsay, France}
\address[Jlab]{Thomas Jefferson National Accelerator Facility, Newport News, VA 230606, USA}
\address[UVA]{Virginia Union University, 1500 N. Lombardy St., Richmond, VA 23220, USA}

\cortext[cor]{Corresponding author}

\date{\today}

\begin{abstract}
A new radial time projection chamber based on Gas Electron Multiplier amplification layers was developed for the BONuS12 experiment in Hall B at Jefferson Lab. This device represents a significant evolutionary development over similar devices constructed for previous experiments, including cylindrical amplification layers constructed from single continuous GEM foils with less than 1\% dead area. Particular attention had been paid to producing excellent geometric uniformity of all electrodes, including the very thin metalized polyester film of the cylindrical cathode. This manuscript describes the design, construction, and performance of this new detector.

\end{abstract}

\maketitle

\textbf{Keywords}: Time projection chamber, gas electron multipliers, spectator-tagging, BONuS12 experiment, Nuclear physics, CLAS12 spectrometer.


\section{\label{sec:intro} Introduction \protect\\}

The BONuS12 experiment was carried out in Hall B at Jefferson Lab in 2020. The goal of the experiment was to measure the $F_2$ structure function from a {\it nearly} free neutron at large Bjorken $x$ via inclusive electron scattering, where $x$ is the fraction of the momentum carried by the struck quark inside the 
nucleon. This experiment utilized the recently upgraded CEBAF accelerator, which nearly doubled the beam energy compared to the previous BONuS experiment ~\cite{PhysRevLett.108.142001, PhysRevC.89.045206}.  This allowed the experiment to reach larger values of both Bjorken $x$ in the deep inelastic scattering (DIS) region, as well as the 4-momentum transfer $Q^{2}$. Due to the lack of high density neutron targets, the experiment utilized a pressurized gas target filled with deuterium gas. In order to ensure that the electron scattered off a weakly-bound neutron inside the deuterium, a low momentum spectator proton was tagged in a Radial Time Projection Chamber (RTPC) in coincidence with the scattered electron. The scattered electrons, as well as other particles, were detected with the CLAS12 spectrometer in Hall B.

 For this RTPC, the gas electron amplification relied on three concentrically stacked layers of Gas Electron Multiplier (GEM) foils ~\cite{Sauli:1997qp}. Its design is a significant improvement over earlier GEM based RTPCs utilized in Hall B for both, the original BONuS experiment~\cite{Fenker:2008zz}, and for a later experiment~\cite{Dupre:2018nim} designed to tag alpha particles. Much of the design improvements were due to progress in the production of large area GEMs, which allowed the production of cylindrical GEM layers made from single GEM foils without the need to splice multiple foils together. This significantly reduced the inactive area of the RTPC from 30\% in the previous RTPCs to 3\% in this unprecedented  detector.

The BONuS12 RTPC replaced the central tracker of the CLAS12 spectrometer~\cite{clas12nim}, which is situated inside the solenoid of the CLAS12 Central Detector. Compared to a more typical ``axial'' TPC, an RTPC has several advantages in view of the needs of this experiment. A larger number of readout pads to improve the spatial resolution and to accommodate larger multiplicity rates. Reduced drift time of the ionization electrons inside the active detector region, making an RTPC a comparatively faster detector. It also allows for a somewhat simpler gas system. 

There were several motivations in designing this new detector: 
\begin{enumerate}
\item Double the detector active length in comparison to the first generation BONuS detector to 40~cm in order to increase the luminosity and counting statistics, and increase the diameter to 160 mm, constrained by room needed for the detector readout and the scintillator detectors installed inside the solenoid magnet,
\item Reduce the size of dead regions in the azimuthal angle $\phi$ to achieve close to $360^{\circ}$ coverage, 
\item Improve the gain uniformity and stability, and 
\item Improve the reconstructed vertex, momentum, and energy loss resolutions. 
\end{enumerate}

In addition to doubling the length, the radial coverage of the drift region was increased from 3~cm to 4~cm to improve the momentum resolution for protons above 70~MeV/c by sampling more ionization points for fitting tracks with a larger radius of curvature. The momentum of the recoiling charged particle is calculated from the radius of curvature of its helical track resulting from the released ionization points within the detector's drift region, which is embedded inside a 4 Tesla solenoidal magnetic field oriented along the beam axis. This is illustrated in the top panel of Fig.~\ref{fig:schematic}, which shows a schematic view of the RTPC looking along the beam axis, as well as a proton track originating from the target and the paths of ionization electrons liberated in the drift gas region as they move under the influence of both the longitudinal $\vec B$ field and the radial $\vec E$ field.  The number of ionization electrons in the charge cloud are amplified as they move radially outward from the drift region through a triple stack of concentric cylindrical GEMs before inducing an electrical signal on a large array of rectangular electrodes on the inner surface of the outer PCB (``padboard'') shell at ground potential.

In addition to the momentum reconstruction of the recoil protons, good resolution of the energy loss per track length ($dE/dx$) is required in order to distinguish protons from heavier nuclear fragments. This is facilitated by a detector that has both stable and uniform pad-to-pad gains. The latter necessitates maintaining good geometrical uniformity of all the electrode layers during detector construction, including the very thin cylindrical cathode foil.

The basic design of the RTPC will be described in the next section. Further details on the RTPC construction will be given in Section~\ref{sec:construction}. Section ~\ref{sec:clas12} will discuss the integration of the RTPC in the CLAS12 spectrometer and the data acquisition, and the calibration for the RTPC will be detailed in Section~\ref{sec:calibration}

A total of three complete RTPC detectors were built for the experiment in order to provide the ability to quickly swap detectors during the experiment in case of a malfunction. This turned out to be 
very useful, as the first RTPC (RTPC-1) did develop a high-voltage connection problem and was exchanged with the third detector (RTPC-3). The latter worked flawlessly for the rest of the data taking. Additional details on the performance of RTPC-3 will be presented in Section~\ref{sec:performance}. 

\section{\label{sec:design} RTPC Design}

\begin{figure}[t]
\centering
    \includegraphics[scale=1.2]{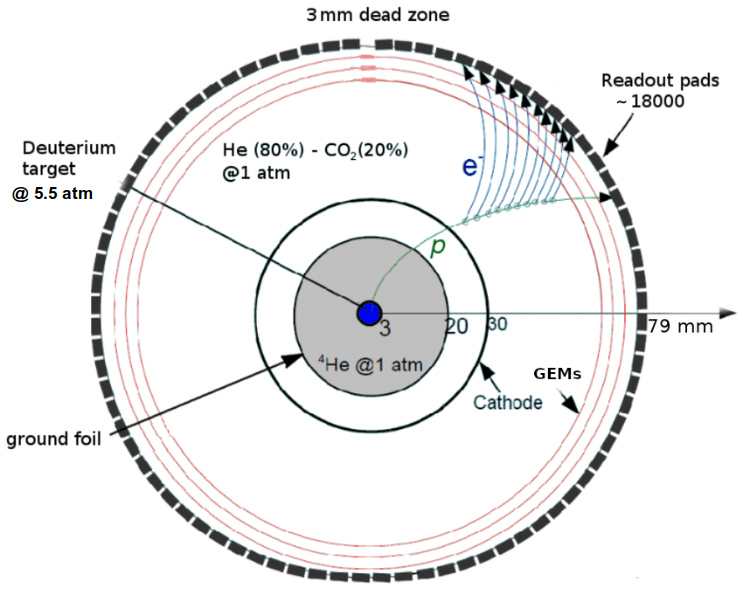} 
    \includegraphics[scale=0.16]{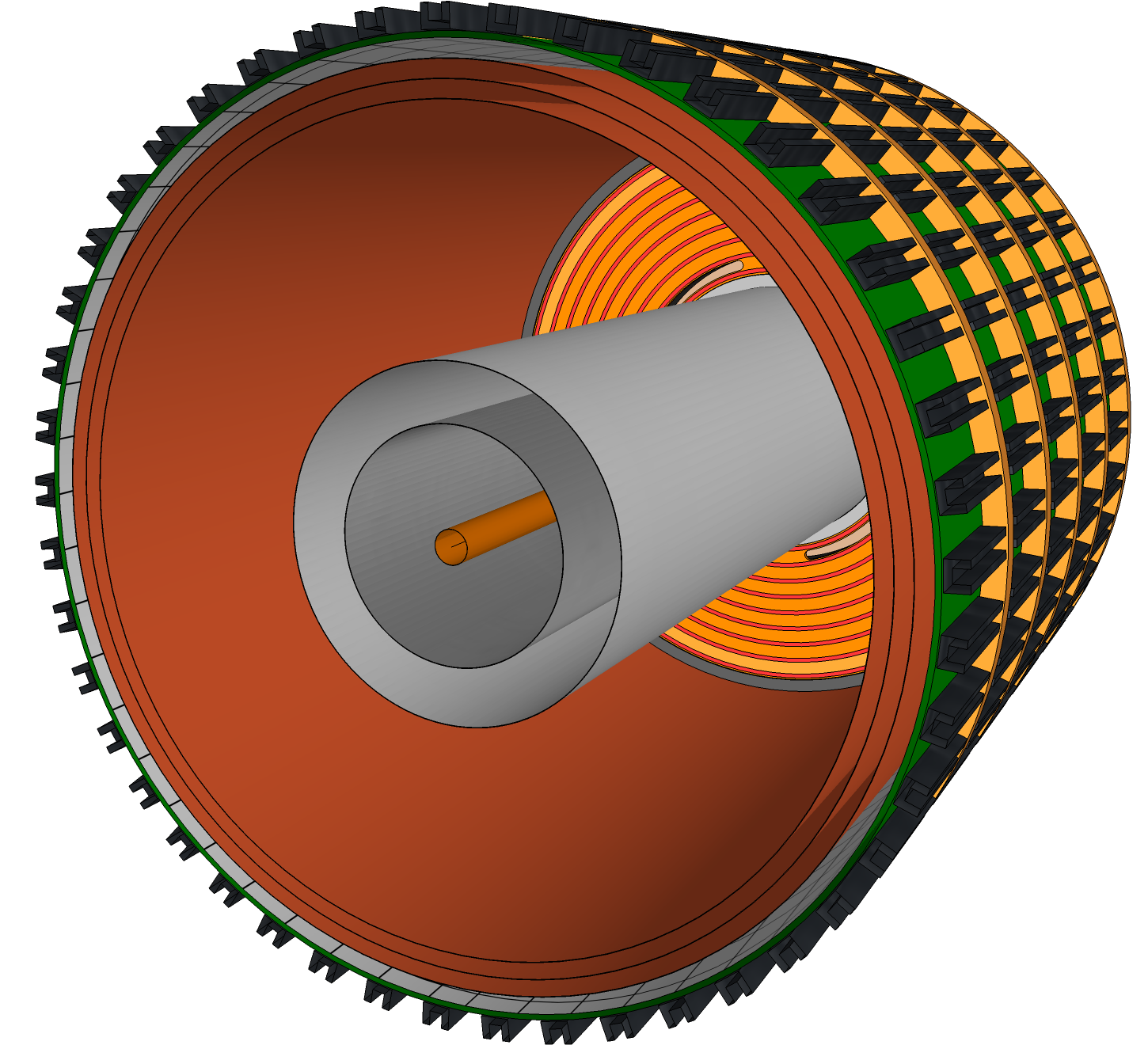}
\caption{Schematic of the BONuS12 RTPC looking along the beam axis (top) and cutaway of the CAD drawing (bottom).}
\label{fig:schematic}
\end{figure}

\begin{figure*}[tbp]
\centering
\includegraphics[scale=0.24] {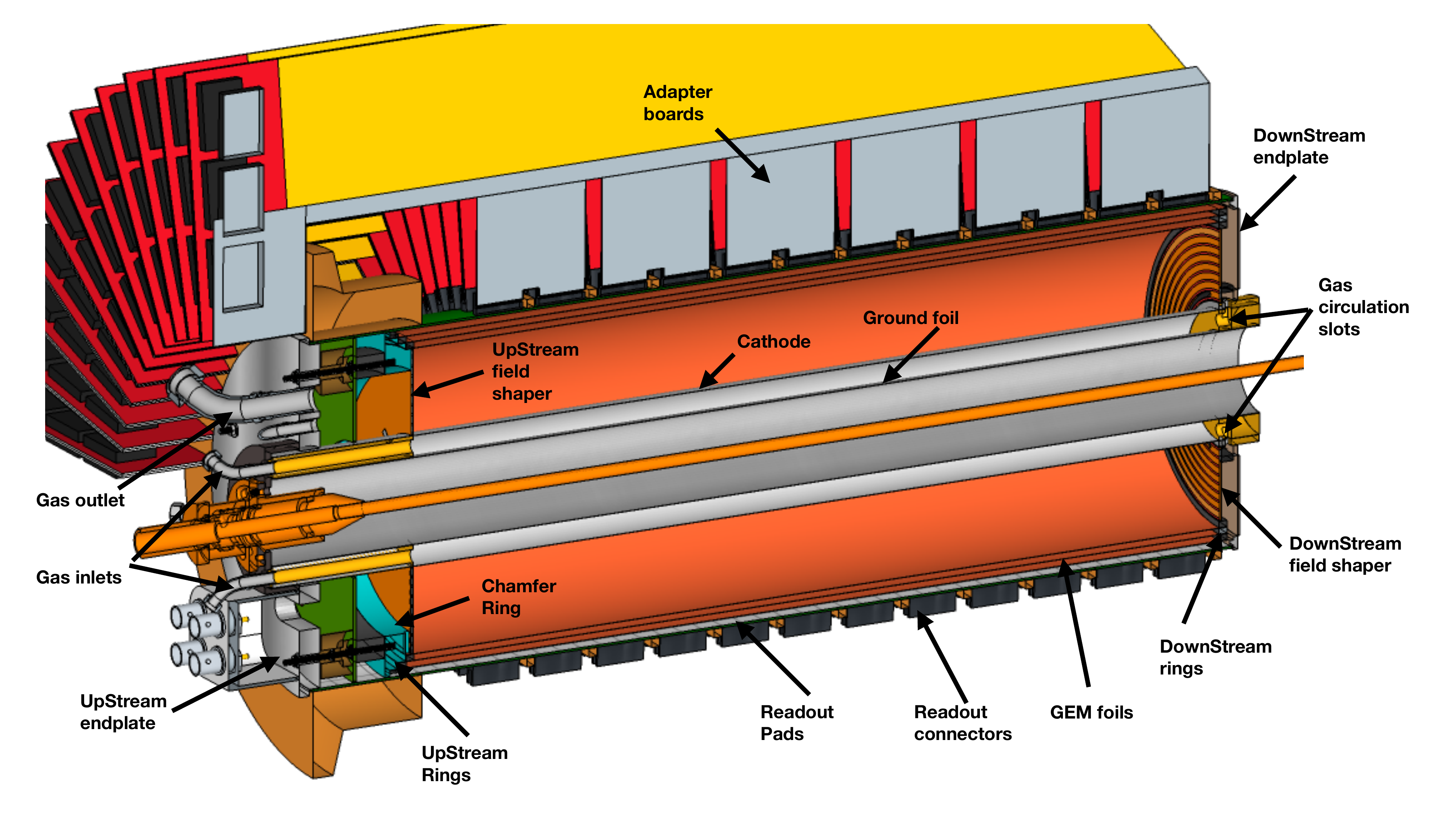}
\caption{CAD drawing of the BONuS12 RTPC showing the major components.}
\label{fig:HW3}
\end{figure*}

 In this section, the different regions and elements of the RTPC, shown in Fig.~\ref{fig:schematic} and Fig.~\ref{fig:HW3}, are presented. Starting from the beamline and moving radially outward, ending with the design specification for the UpStream (US) and DownStream (DS) end plates, the RTPC is composed of: 

\begin{itemize}

\item \underline{Target:} 
The target straw is made of a spiral-wound aluminized Kapton (polyimide) foil with 63$~\mu m$ wall thickness and 6~mm diameter, maintained during data taking at a pressure of up to 5.5 atmospheres. The straw is approximately 50 cm long (extending beyond the length of the RTPC on both ends, with both entrance and exit windows outside the RTPC acceptance). It is aluminized to minimize leakage and the Aluminum coating thickness is 0.1$~\mu$m.

\item \underline{Buffer region:}
The region between the outer target straw radius and the ground foil (see below). It is filled with gaseous $^4$He at atmospheric pressure to minimize secondary interactions from M\o{}ller electrons produced by the electron beam. This same volume extends beyond the DownStream end of the RTPC in the form of a Kapton foil ``snout'', up to the entrance opening of the M\o{}ller shield, to minimize background and scattered electron rates from the beam outside the RTPC fiducial acceptance.

\item \underline{Ground foil:} The ground foil is a cylinder made from a 6~$\mu$m thick aluminized mylar foil and positioned co-axially at 20 mm from the center of the target straw. The purpose of the ground foil is to prevent the target surface from collecting charges due to the field generated by the cathode.  All volumes from the ground foil outward to the padboard are filled  with the same drift gas, a mixture of 80\% $^4$He and 20\% CO$_2$ gas by volume.

\item \underline{Cathode:}  The cathode cylinder is made from a 6~$\mu$m thick aluminized mylar foil and positioned at 30~mm radial distance from the center of the beamline. Under standard operating conditions it was held at 6.5~kV relative to the ground and the outer readout anode layer. A CAD drawing of the cathode is given in Fig.~\ref{fig:cathode}. The rings labeled US ring and DS outer ring are made of polyetherimide plastic, while the DS inner ring is made of Rohacell foam. The ground foil is attached via epoxy to the inner surface of these rings, while the cathode is attached to the outer surface, and the DS outer ring is epoxied to the outer cathode foil surface. The UpStream (US) ring has four through holes equally spaced to which the drift gas inlet ports are epoxied. The notches in the DownStream (DS) inner foam ring and the four equally spaced through holes in the DS outer ring provide a path for the gas to flow radially outward to the DownStream region beyond the field shaper, which effectively acts as a small gas buffer region.  

\begin{figure}[h!]
\centering
\includegraphics [width=0.5\textwidth,height=0.35\textwidth] {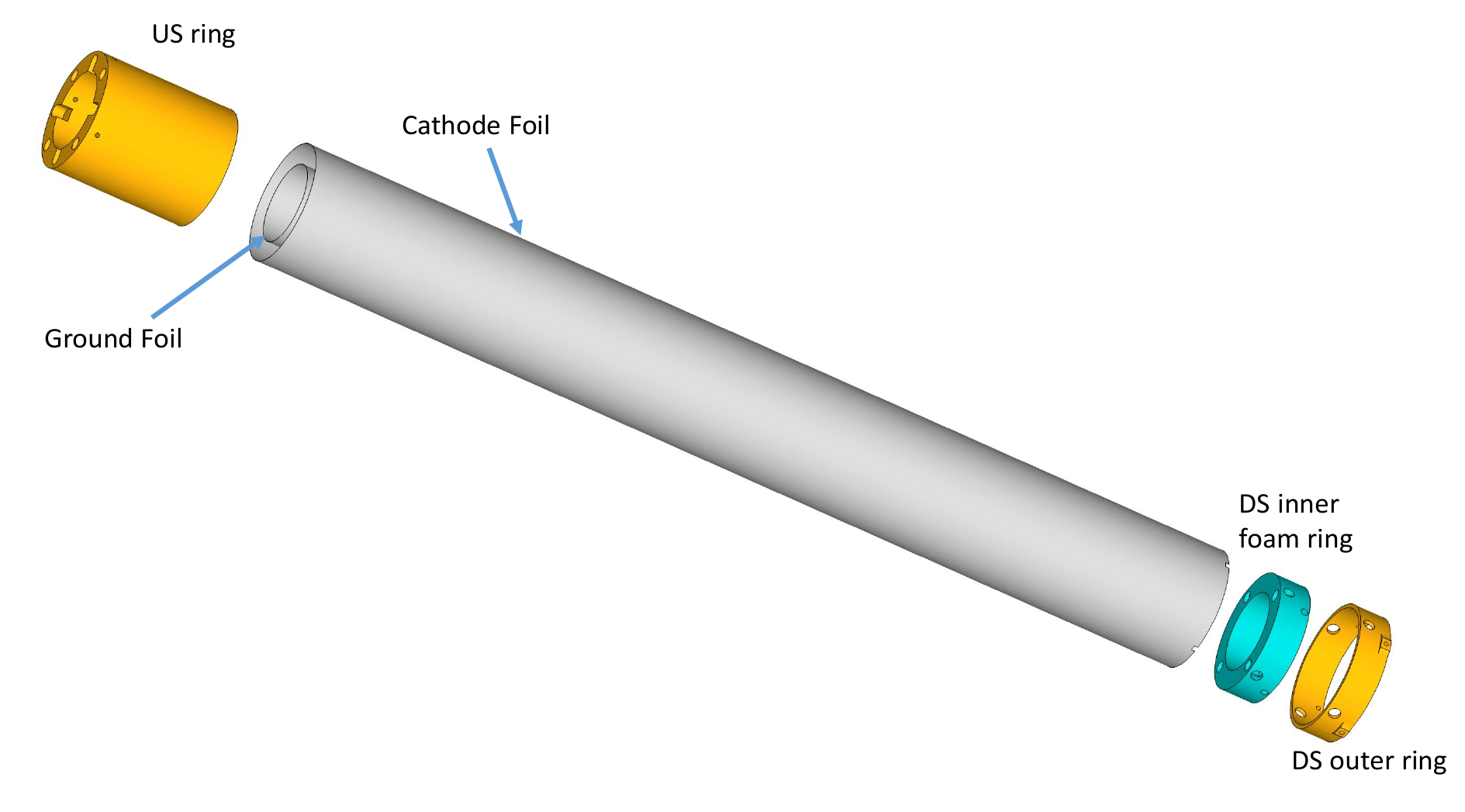}
\caption{CAD drawing of the cathode assembly and an exploded view of its individual rings.}
\label{fig:cathode}
\end{figure}   

\item \underline{Drift region:} This region extends from the cathode to the first GEM, 70~mm away from the beam axis.  
      The electric field in this region is radial and averages around 1100~V/cm.

\item \underline{Electron amplification system:} This system consists of three cylindrical GEM foils located at inner radii 70, 73, and 76~mm. The 50~$\mu$m thick GEM foils utilized were produced at CERN (European Organization for Nuclear Research) and had a double conical hole shape with standard hole radii of 70~$\mu$m (outer) and 50~$\mu$m (inner) and a standard pitch of 140~$\mu$m. This system amplifies the charge collected by the readout channels via gas amplification. All GEM foil cylinders have a similar design with an epoxy seam 
parallel to the detector axis, consisting of a 3~mm wide dead region with no copper coating or amplification holes. Each GEM foil is segmented into sixteen high-voltage sectors in the azimuthal direction in order to limit the energy released from possible discharge events. The detailed dimensions of the different GEM foils are presented in Table~\ref{tab:tableHW1}. The radial precision and uniformity along the central axis of the detector ($z$) of each GEM foil is about 100~$\mu$m, which corresponds to 3.3\% of the radial separation between the GEM foils. The first GEM layer is set to $\Delta$V= -2025~V relative to the ground and then each subsequent layer is set to a lower voltage relative to the previous to obtain a strong (1000 V/cm) electric field between the GEM foils. A 370 V bias is applied across each GEM for amplification.

\begin{table}
\centering
\begin{tabular}{lccc} 
\Xhline{1pt}
& Inner & Middle & Outer \\ [0.5ex] 
& (mm) & (mm) & (mm)\\
 \hline
 Radius & 70  & 73  & 76 \\ 
   \begin{tabular}{@{}l@{}}Width \\ (outer radius circumference )\end{tabular}  & 440  & 459  & 477.8 \\
 Total width  including overlap & 443 & 462.13  & 481.05 \\
 Active area width & 437  & 455.87  & 474.55  \\
 Active area length & 400  & 400  & 400 \\
 DownStream inactive length & 4 & 4  & 4 \\
 UpStream inactive length & 4  & 8  & 12 \\
 Sector width & 27.125  & 28.3  & 29.47 \\
 Sector length & 400  & 400 & 400 \\
 Sector clearance & 0.2  & 0.2  & 0.2 \\
 Power strip length & 30  & 45  & 60 \\
 Seam width & 3  & 3.13  & 3.25 \\
\Xhline{1pt}
 \end{tabular}
 \caption{\label{tab:tableHW1} GEM foil dimensions~\cite{Aruni2022}}.
  \end{table}
   
\item \underline{Readout Padboard:} This board is positioned at an internal radius of 79~mm and collects charges after they get amplified by the GEMs. The inner surface of the padboard is segmented into 17,280 copper pads including dead-space between pads for isolation. Each pad covers 3.9~mm in $z$ and 2$^{\circ}$ in $\phi$, while the dead-space between pads is about 0.1~mm. Each row of pads, within a set of four rows, is staggered by 1~mm to improve the spatial resolution. A portion of the layout is shown in Fig.~\ref{fig:pads}. The flexible padboard is made of FR4. The copper clad readout pads on the inside were connected to the external connectors by vias. The connectors are soldered on the outside surface and connected to flexible signal translator boards (see Fig. 2). 

\begin{figure}[h!]
\centering
\includegraphics [scale=0.23] {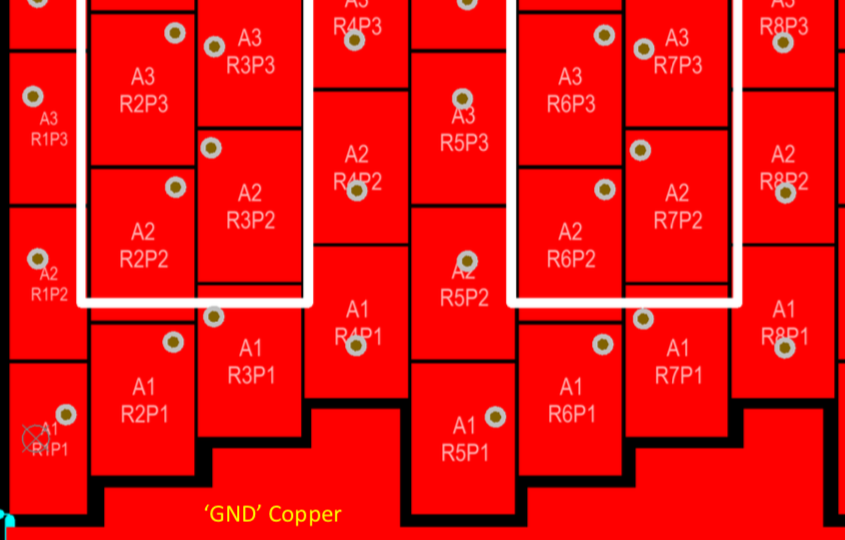}
\caption{Layout of the cylindrical readout pad plane. The $z$ beam axis is pointing up-down. The individual pad size is 2.7~mm perpendicular and 3.9~mm parallel to $z$. Note that each row is staggered by 1~mm with respect to the previous one, for each set of 4 rows.}
\label{fig:pads}
\end{figure}

\begin{figure*}
\centering
\includegraphics[scale=0.22]{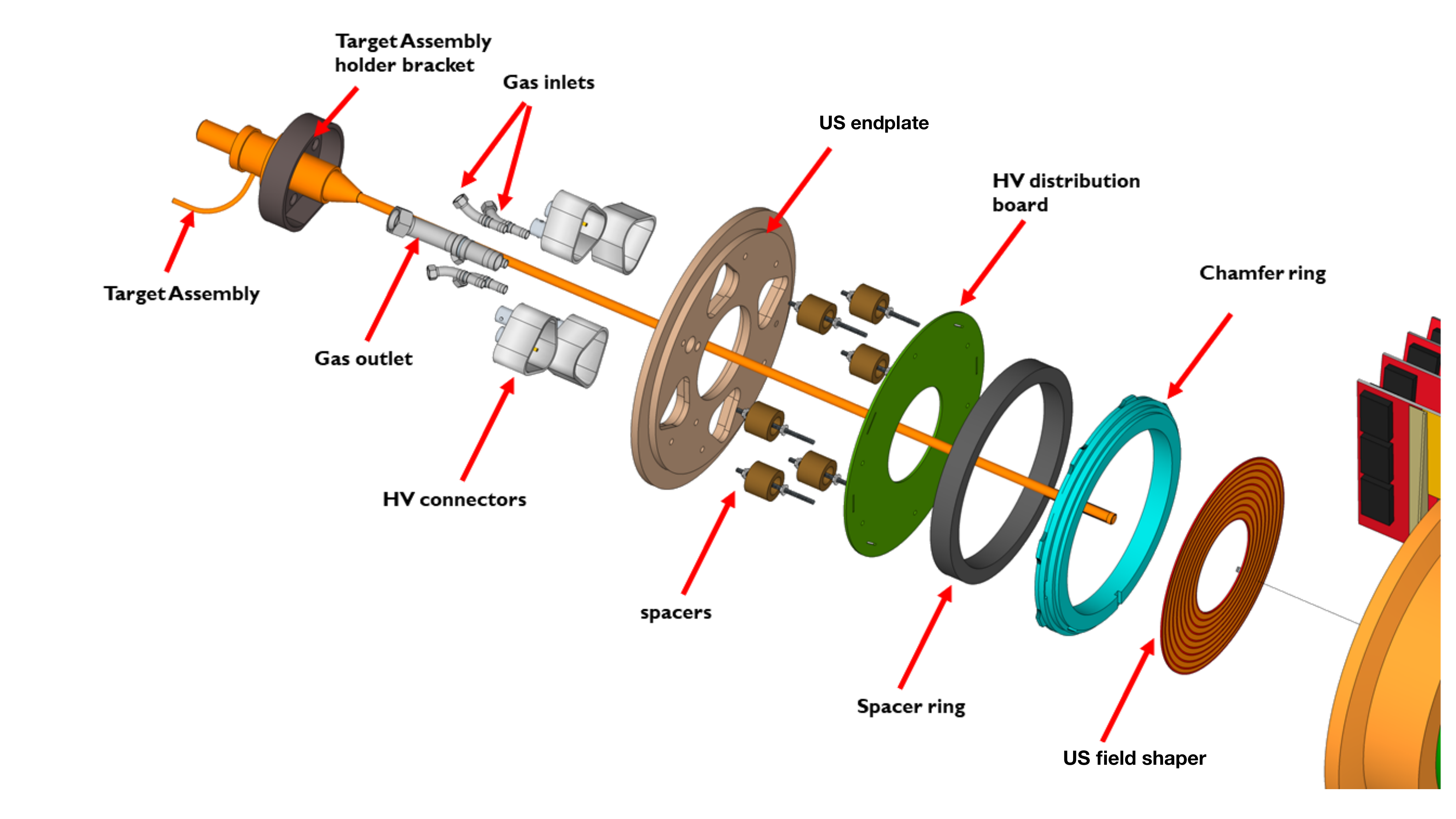}
\caption{Exploded CAD view of the UpStream (US) assembly. From left to right, the individual elements are: Target straw and fill assembly, Target assembly holder bracket, Gas inlets and outlet, HV connectors, fiberglass US end plate, Spacers and assembly bolts, HV distribution board, Delrin Spacer Ring, Chamfer Ring, and the US field shaper printed circuit boards.}
\label{fig:upstream-assembly}
\end{figure*}

\begin{figure}
\centering
\includegraphics[scale=0.13]{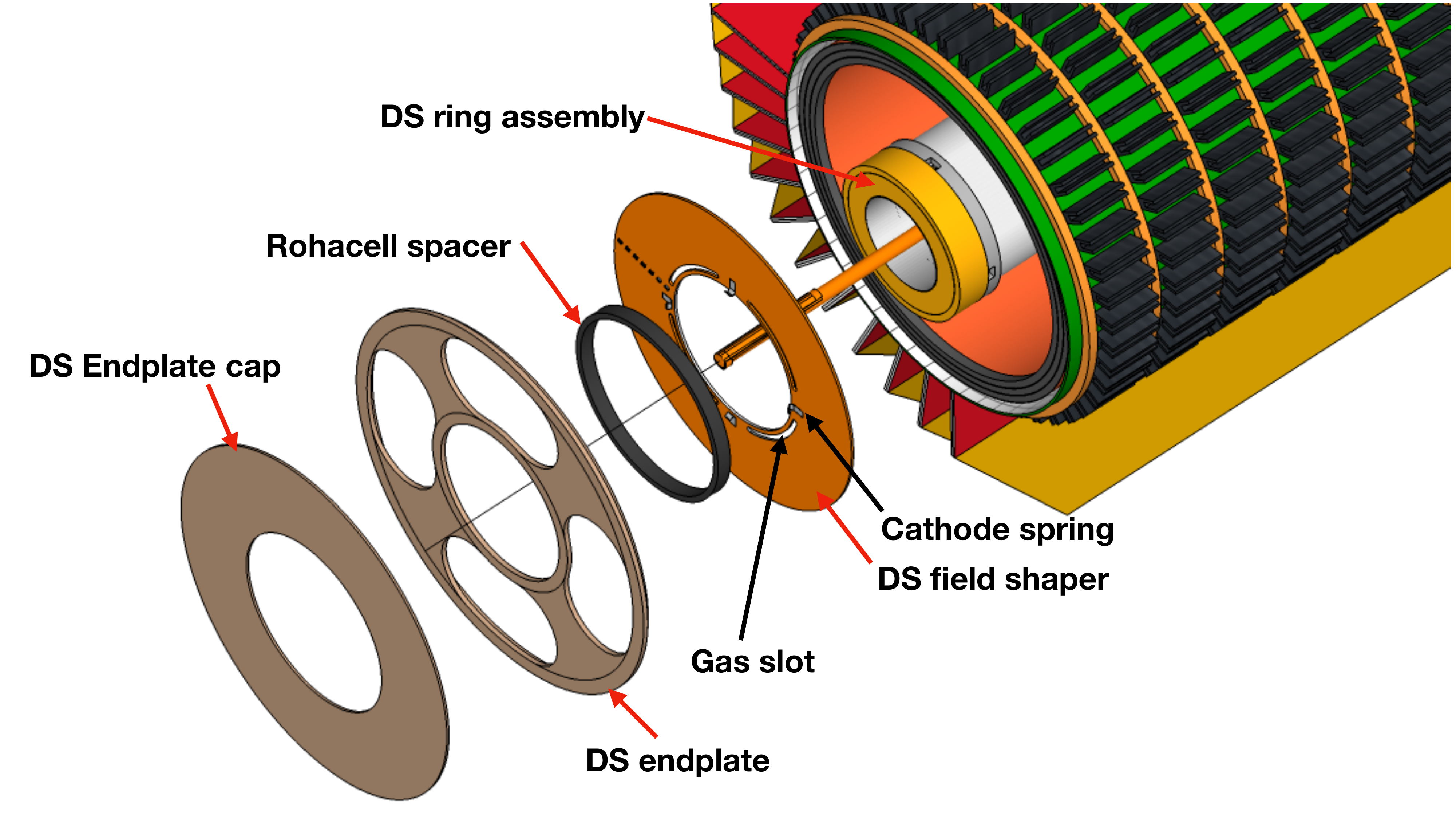}
\caption{Exploded CAD view of DownStream (DS) assembly. From left to right, the individual elements are: Endplate cap, DS endplate, Rohacell spacer, and the DS field shaper.}
\label{fig:downstream-assembly}
\end{figure}

\item \underline{The UpStream (US)}: The assembly, shown in Fig.~\ref{fig:upstream-assembly}, consists of the following elements: the US end plate, Spacers, 
the High Voltage (HV) distribution board, a Spacer Ring, the Chamfer Ring, and the US field shaper. The HV distribution board consists of \SI{10}{\mega\ohm} current limiting resistors. The high voltage tabs from the GEM foils were inserted into the slots placed in the board and soldered into the board. 

The Chamfer Ring is designed as a supportive structure to hold the GEM foils and to provide the axial alignment of the GEMs. It is a stepped ring as shown in Fig.~\ref{fig:upstream-assembly} with dimensions as given in Table~\ref{tab:chamferdim}. The inner radial surface of each step provides the axial alignment for each GEM layer. Each step is beveled such that the radius is about 100~$\mu$m smaller on the DownStream edge and then flares to the proper alignment radius towards the UpStream base. This allows for the GEM inner rings to slide easily onto each step, while still providing plenty of surface for proper alignment.

\item \underline{The DownStream (DS)}: The assembly, shown in Fig.~\ref{fig:downstream-assembly}, consists of the following elements: the DS end plate, the DS end plate cap, a Rohacell spacer, and the DS field shaper. The latter has four gas slots cut through the PCB spaced equally around the azimuth near the inner radius. The RTPC drift gas flows between the ground and cathode foils from the inlets in the US ring and exits radially outward into a DS buffer region through holes in the DS ring assembly before flowing into the drift region through the DS field shaper slots.  Four soldered, metallic spring contacts mounted at the inner radius slide over the cathode foil on the surface of the DS ring assembly to provide electrical connection between the cathode and the field shaper resistor chain.  

\begin{table}
\centering
\begin{tabular}{lcc} 
\Xhline{1pt}
  &  Radius & Length in $\hat z$ \\ [0.5ex]
  & (mm) & (mm) \\
\hline
Inner Radius & 60.0 & 13.0 \\ 
Inner GEM Step & 69.9 & 9.0 \\ 
Middle GEM Step & 72.9 & 5.0 \\
Outer GEM Step & 75.9 & 4 \\
Outer Radius & 78.9 & 2.5 \\
 \Xhline{1pt}
 \end{tabular}
 \caption{\label{tab:chamferdim} Chamfer Ring dimensions}
 \end{table}

The two identical field shapers at the UpStream and DownStream ends of the detector are designed to maintain the uniformity of the electric field inside the drift region. They are made out of printed circuit boards (PCBs) with voltage divider resistors, each of $\SI{500}{\mega\ohm}$. The US field shaper was epoxied to the DownStream side of the Chamfer Ring after alignment with a set of alignment pins. A set of four assembly bolts connect the US end plate to the Chamfer Ring after passing through a set of through holes in the intervening material. The UpStream edge of the readout padboard was epoxied to the stepped edge routed into the UpStream plate at a radius of 78.9~mm, which provide both alignment and a seal for the UpStream gas buffer region. 

\end{itemize}

\section{\label{sec:construction} Construction}

\subsection{\label{sec:gems-test} GEM Quality Controls and Selection} 
Three sets of GEM foils of different surface area were used for the inner, middle, and outer foil cylinders. Seven foils of each size were manufactured by CERN. A set of criteria and tests were established to evaluate the quality of the foils. These criteria were: 
\begin{itemize}
   \item \underline{Optical inspection:} The surfaces of each GEM foil were optically inspected for any outliers. A digital microscope was used to scan the surface of the GEM foils after flushing with nitrogen gas to remove dust and other contaminants. The foils were classified according to the quality of their surfaces. Fig.~\ref{fig:gem_optical_inspection} shows examples of GEM foils considered to pass (top) or to fail (bottom) the optical inspection. 

\begin{figure}[t]
\centering
    \includegraphics[scale=0.16]{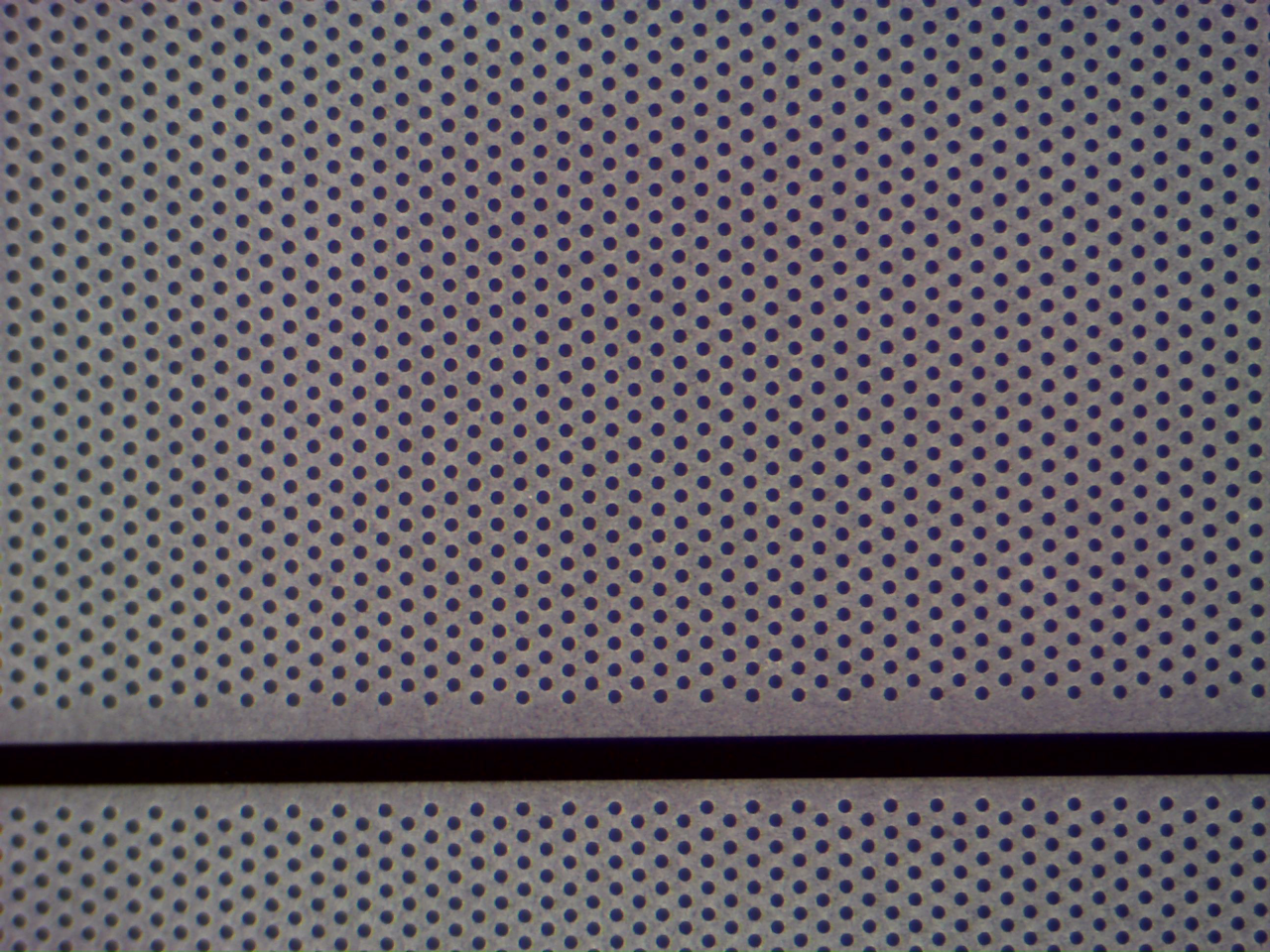} 
    \includegraphics[scale=0.16]{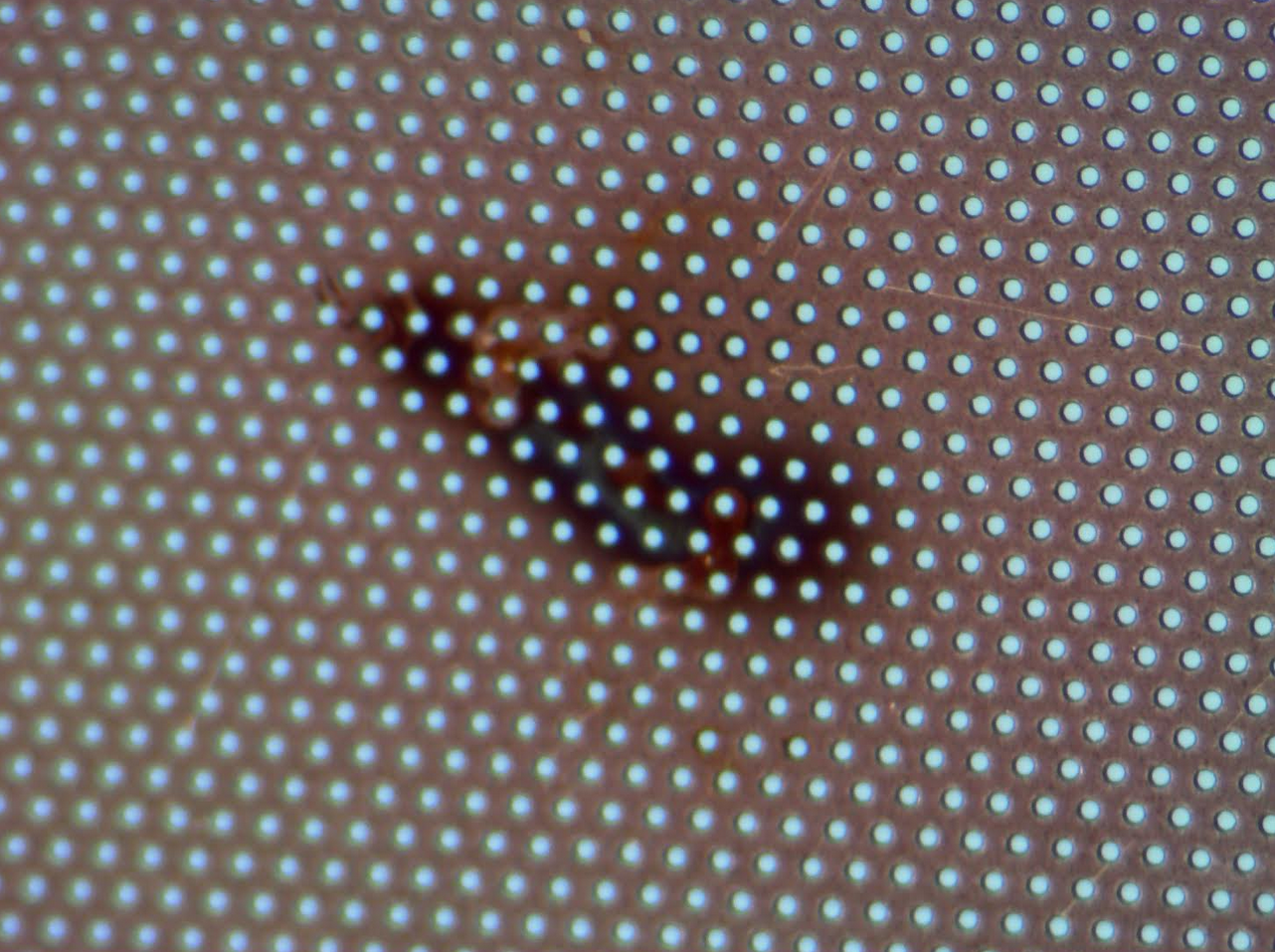}
    \caption{Images from the optical inspection of the GEM foils. The top image shows a good GEM foil that passed the selection criteria for usage in the RTPC. The bottom image shows a GEM foil with some surface deformations and non-perfect hole shapes through the GEM foil, removing this GEM foil from our list of the good foils.}
    \label{fig:gem_optical_inspection}
\end{figure} 

    \item \underline{Continuity of sectors:} Testing the electrical continuity of the HV connection tabs for each of the 16 individual sectors. The continuity test is to ensure the individual connection tabs are properly connected to the respective sector of the GEM foil and that there are no connections between different sectors.

  \item \underline{HV testing for the GEM foils:} During experimental operation, the GEM foils needed to hold 370-390~V continuously without tripping.  During testing, they were required to hold 500~V in order to be considered for installation in the final detectors. This test was performed in a clean box with nitrogen flowing through the box which was pressurized to 10~psi. Foils with leakage currents less than 3~nA passed this selection criterion. 

\end{itemize} 

\subsection{\label{sec:gems-assembly} Forming and Stacking of Cylindrical GEMs} 
\begin{figure}
\centering
\includegraphics[scale=0.11]{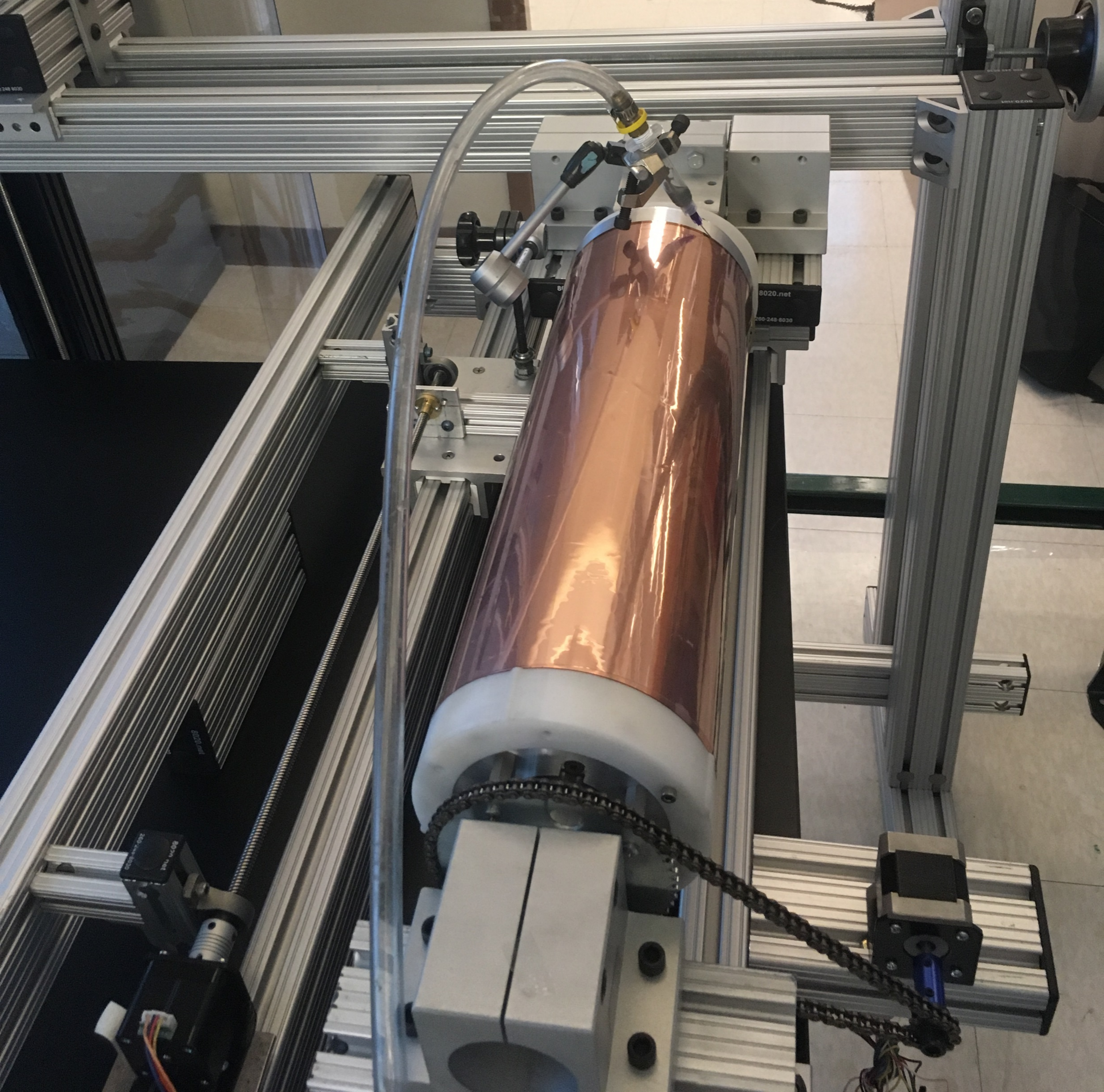}
\caption{Picture of wrapped inner GEM foil on mandrel during application of epoxy for the UpStream ring.}
\label{fig:HW6}
\end{figure}

Cylindrical mandrels were designed and machined from aluminum in order to provide a surface upon which to form the GEM foils into cylindrical shapes. In the wrapping station, the inner GEM foil was first stretched by hand on the mandrel. Under full tension, the inactive overlap regions without copper were epoxied to produce a 3~mm seam of overlapping Kapton along the cylindrical axis (see Fig.~\ref{fig:HW6}). The epoxied foil was then wrapped with a temporary cover sheet of thick Mylar that was tensioned by hand and taped into place to provide registration during overnight curing. Next, the UpStream and DownStream rings were epoxied to the inner and outer surfaces, respectively. The mandrel was milled on the UpStream end, such that the outer surface of the inner ring was flush with the mandrel surface supporting the GEM foil.  After curing over night, the mandrel was then oriented vertically to accommodate the removal of the completed GEM cylinder. The latter was done using an actuator puller assembly attached to the DownStream ring. Next, the GEM foil was removed from the mandrel using the puller assembly.

\begin{figure}
\centering
\includegraphics [width=0.48\textwidth,height=0.40\textwidth] {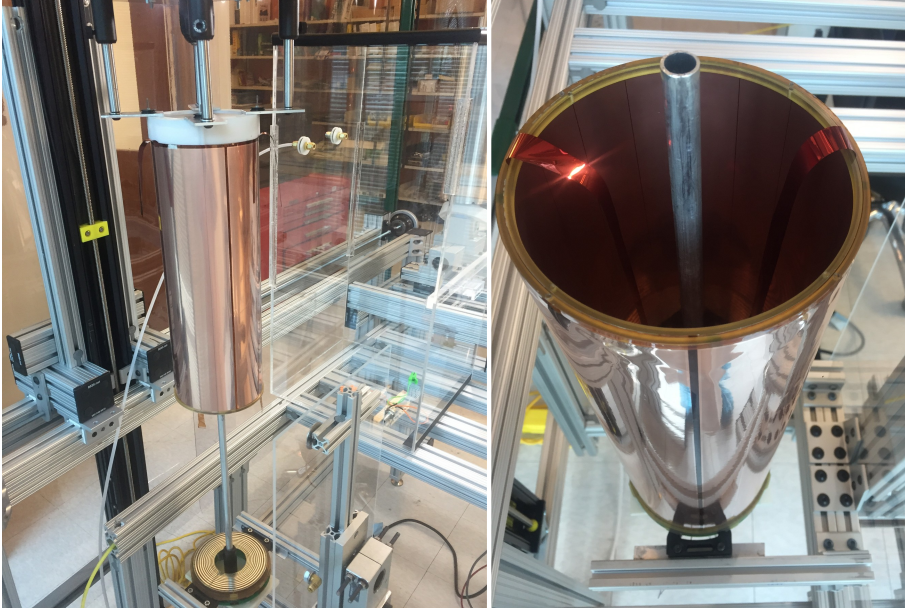}
\caption{Various stages of the GEM stacking process. (Left) Inner GEM being lowered onto chamfered ring utilizing a linear actuator and mating self-alignment posts. (Right) View of the stacked triple GEM layers from the DownStream end.  Visible are the DownStream spacer rings and the HV tabs for the connection to the outer ring of the DownStream field shaper.}
\label{fig:gem-install}
\end{figure}
Subsequently, the GEM was moved inside an HV test station and tested up to 400~V between the inner and outer sides of the GEM foil. In case of a successful HV test, the GEM foil was installed on the Chamfer at the final assembly station as shown in Fig.~\ref{fig:gem-install}. The process was repeated for the remaining middle and the outer GEM foils on the final assembly station from inside to outside to complete the cylindrical triple GEMs. All the GEMs were installed on the Chamfer with the help of a puller assembly with a coaxial guiding rod centered as shown in Fig.~\ref{fig:gem-install}.

\subsection{\label{sec:padboard-assembly} Readout Padboard} 
\begin{figure}[h!]
\centering
\includegraphics [width=0.23\textwidth,height=0.45\textwidth] {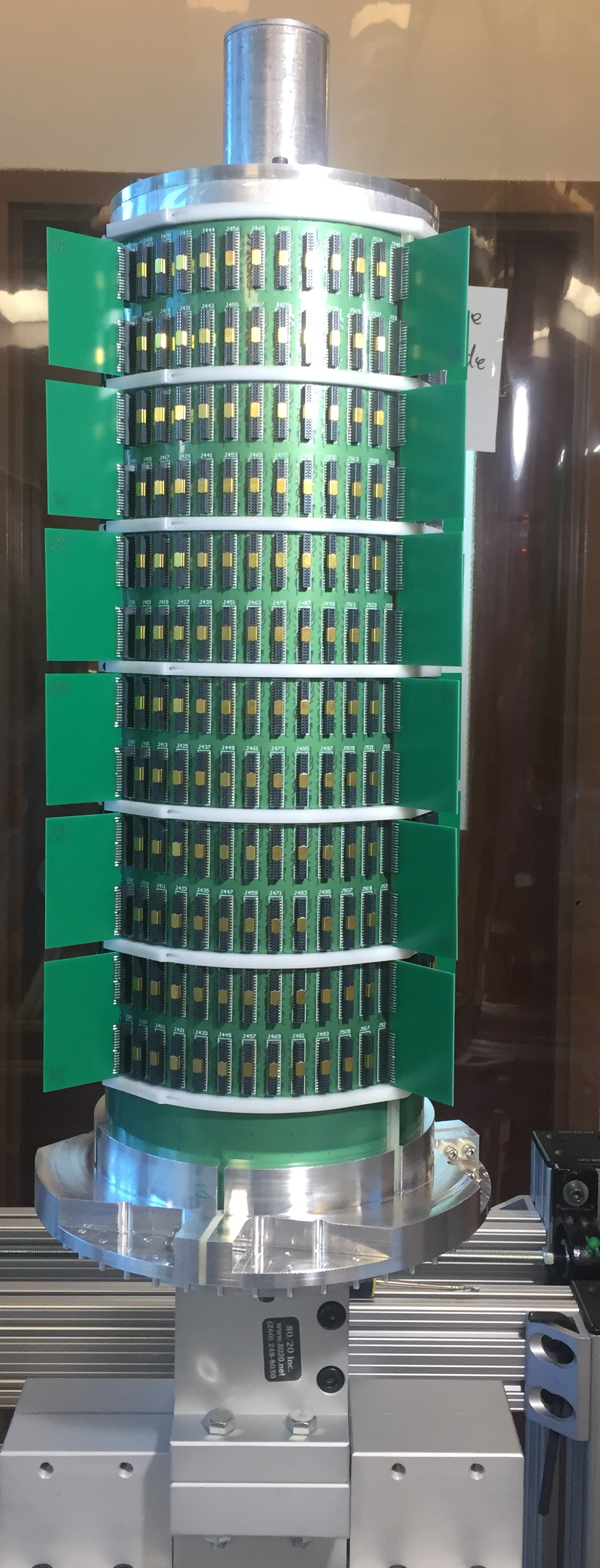}
\includegraphics [width=0.23\textwidth,height=0.45\textwidth] {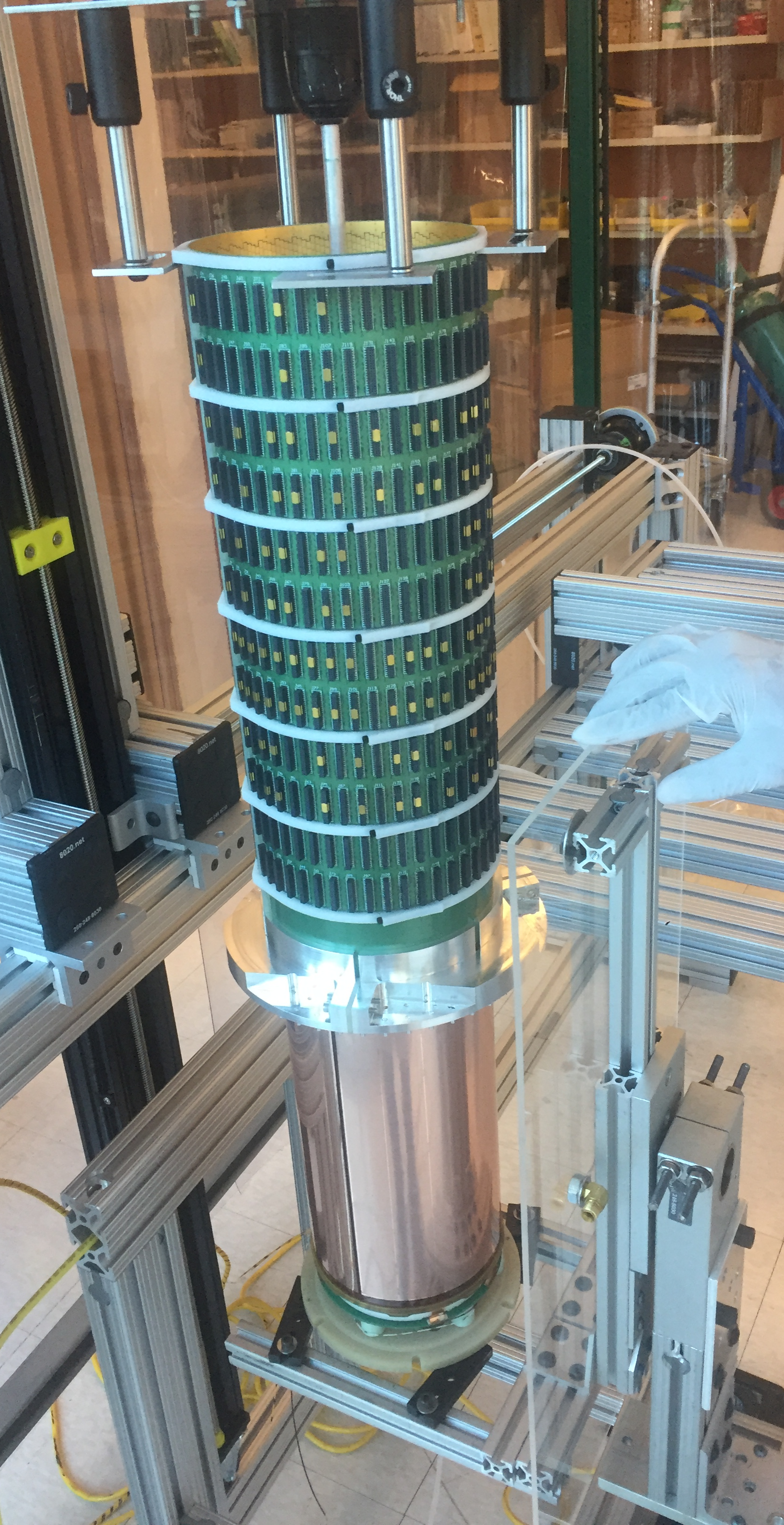}
\caption{(Left) Picture of the padboard after wrapping it on the mandrel, where some cards are used to align the rings on the surface. (Right) Picture of the padboard installation on the final assembly station.}
\label{fig:padboard-install}
\end{figure}

\begin{figure}[h!]
\centering
\includegraphics [scale=0.08] {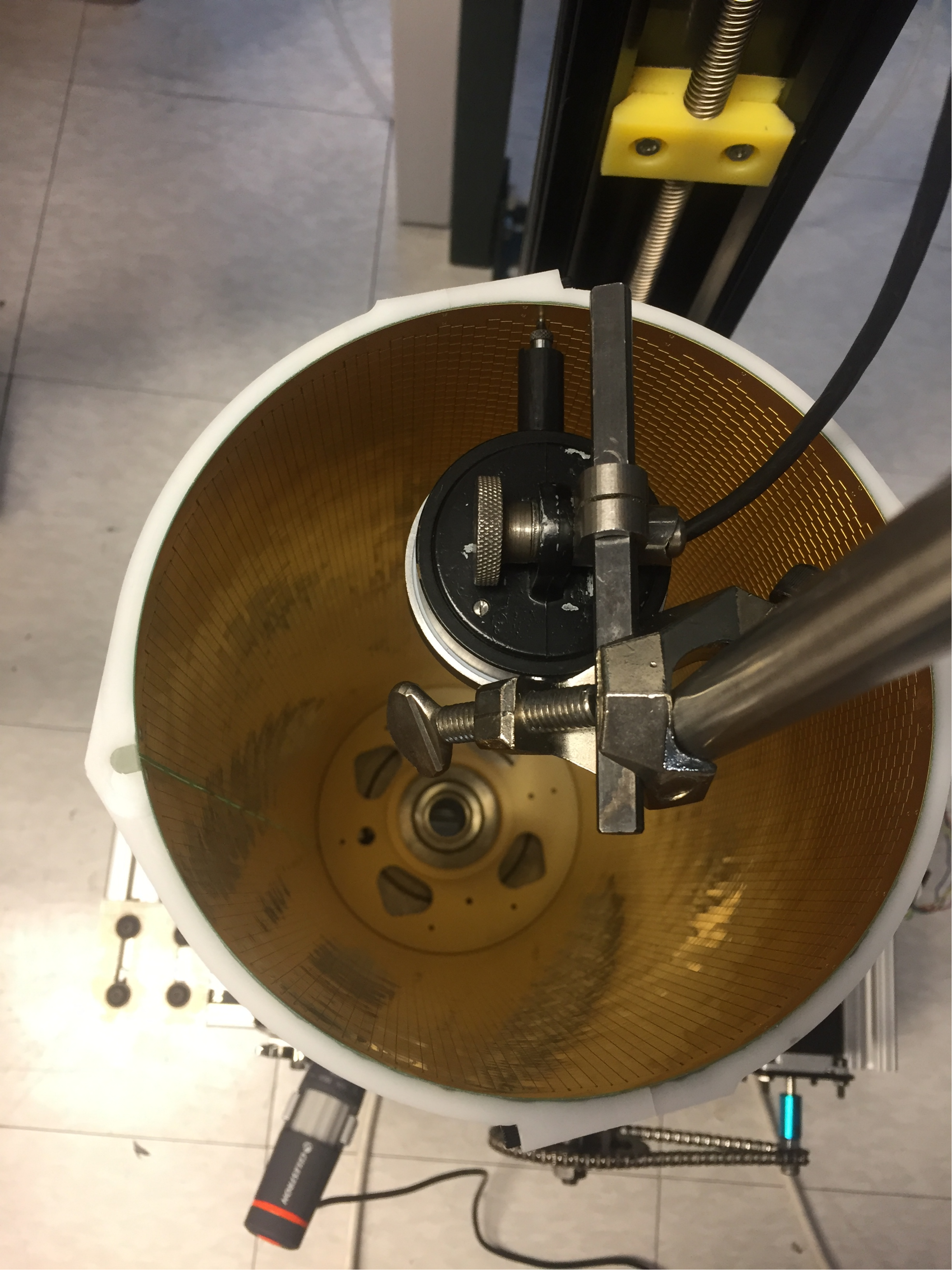}
\includegraphics [scale=0.5] {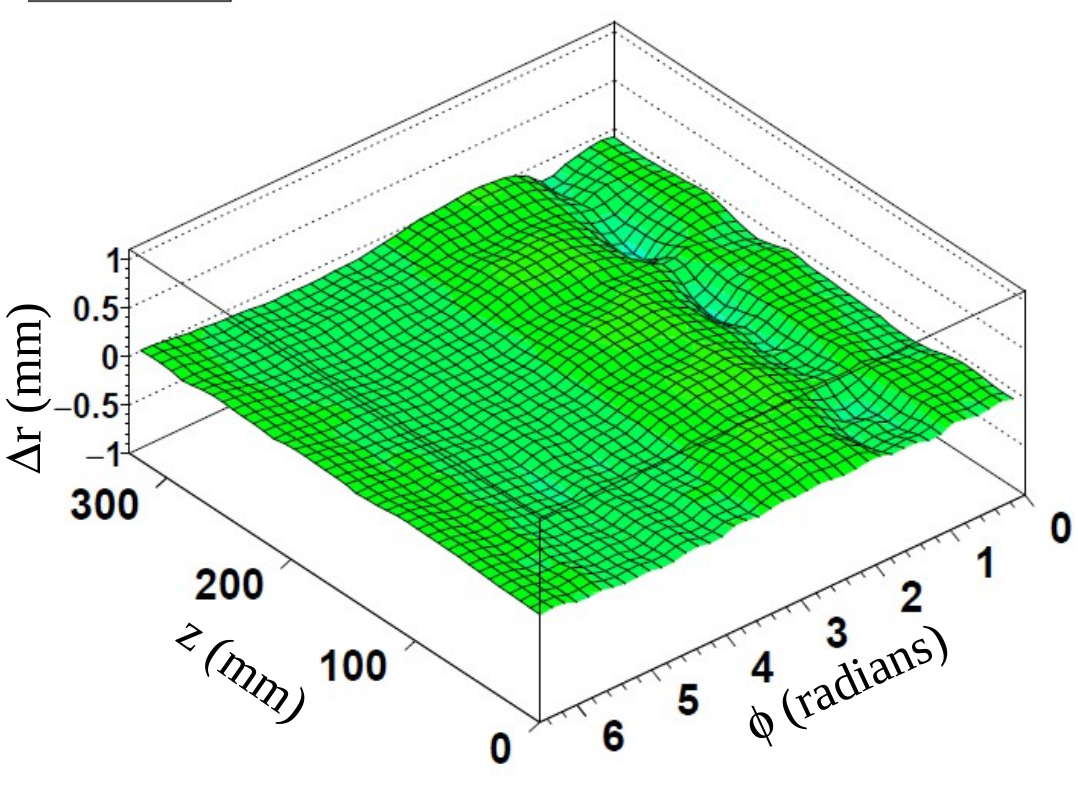}
\caption{(Top) Picture of the padboard mounted on the automated surface scanner. (Bottom) Results from the analysis of the surface scanner showing variations in the surface radius, given in mm as a function of $z$ and $\phi$} 
\label{fig:padboard-uniformity}
\end{figure}

The readout padboard was wrapped around its mandrel and held in shape by six sets of rings, as shown in Fig.~\ref{fig:padboard-install}. Then, a heavy ring was glued to the DownStream edge and the spine, and a 3~mm thick square fiber glass rod was glued on to the joint of the two edges of the padboard. After keeping it for sufficient time to cure the glue on the mandrel, the padboard was removed from the mandrel. The padboard surface uniformity was measured using a cylindrical surface scanner developed at Hampton University. A picture of the padboard mounted on the automated scanner is shown in the top panel of Fig.~\ref{fig:padboard-uniformity}. The scanner consisted of an electronically recorded depth gauge measuring the radial variations of the padboard inner surface, a linear actuator for stepping the gauge along the axis of the cylinder, and a turntable connected to a stepper motor for rotation in small increments in $\phi$.  Although care was taken to align the axis of the cylinder with that of the gauge arm, any remaining misalignments produce a characteristic sinusoidal behavior of the measured variation and were subsequently fitted and corrected for when determining the true radial variation. The measured radial variation, in mm, is shown versus $z$ and $\phi$ in the bottom panel of Fig.~\ref{fig:padboard-uniformity}. These were measured to typically be less than 150~$\mu$m, with the largest variations of $\approx$ 300~$\mu$m measured near the seam (seen at a little less than 1~$\pi$ radians in the figure).

Subsequently, an electrical test was performed to ensure that no short circuits existed between readout pads. In the first detector's padboard, about 2\% of the pads were found with open circuits, which were fixed by re-soldering the faulty connector pins. 
Finally, the padboard was installed on the final assembly station, as shown in Fig.~\ref{fig:padboard-install}, and the UpStream aluminium end plate was used to attach the padboard to the detector assembly.

\subsection{\label{sec:cathode-assembly} Construction and Installation of the Cathode Assembly } 
The cathode and the ground cylinders were made from aluminized mylar foil of 6~$\mu$m thickness. Such a thin foil is difficult to work with during both the cutting and wrapping process without additional forces to keep the foil surface wrinkle free.  During the cutting process an electrostatic method was used for this purpose.  The foil was first laid out onto a thin plastic sheet, which sat on a layer of Aluminum foil.  A power supply limited to low current provided approximately 200~V potential difference between the foil layers and an electric force on the thin mylar foil directed towards the plastic surface.  This allowed the foil to be pulled flat against the surface and to remain so while cutting.  In order to keep the foil free of wrinkles during wrapping, an inward radial force against the mandrel surface was applied via air pressure.  The mandrel consisted of a hollow nylon cylinder with small radial holes through the cylindrical shell, uniformly spaced along the surface.  A small vacuum pump was connected to the end of the cavity to provide the pressure differential keeping the foil against the mandrel surface.  This kept the foil in place while smoothing out any wrinkles prior to applying epoxy.

\begin{figure}[h!]
\centering
\includegraphics [width=0.23\textwidth,height=0.45\textwidth] {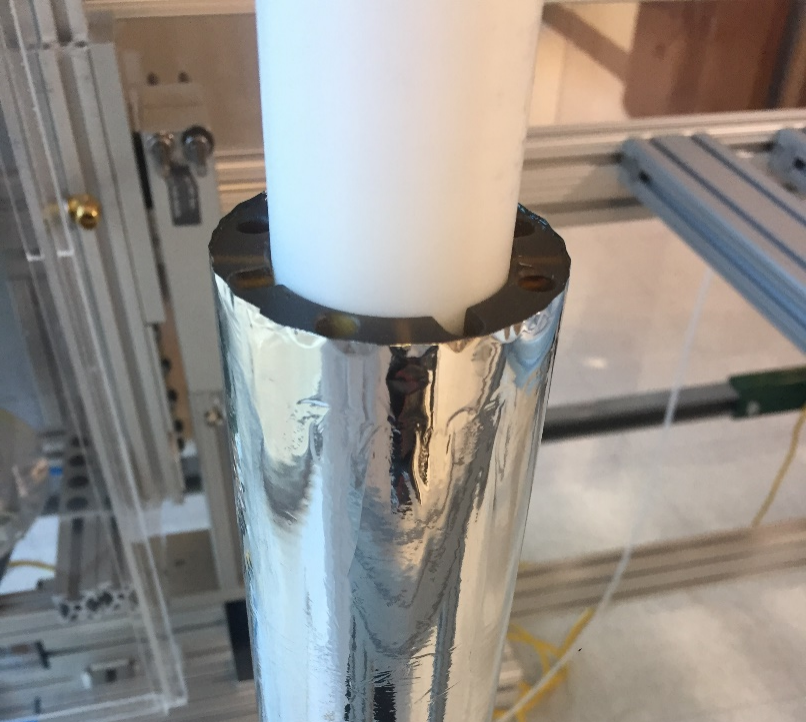}
\includegraphics [width=0.23\textwidth,height=0.45\textwidth] {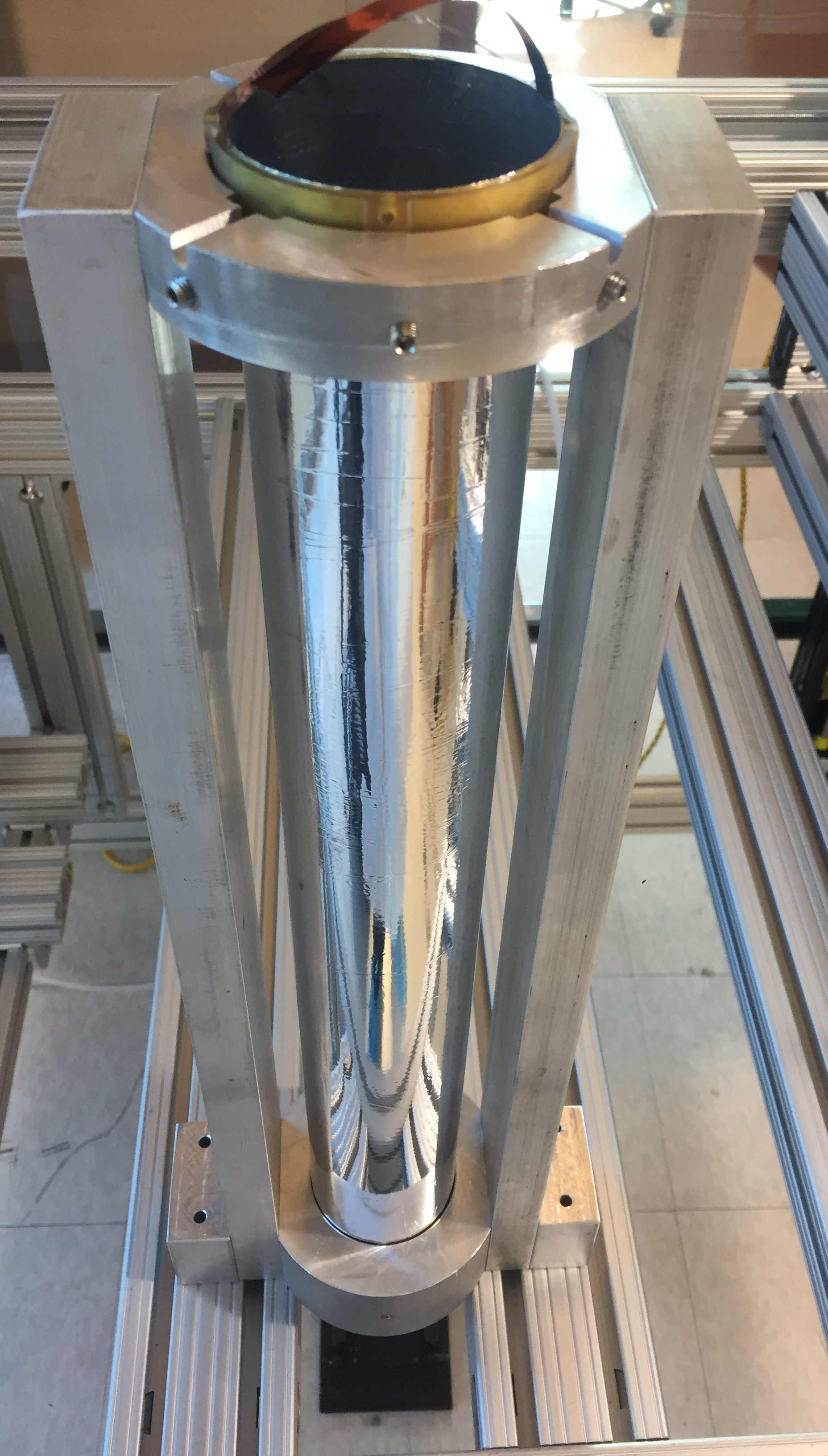}
\caption{(Left) Picture of the UpStream side of the cathode foil on the mandrel. (Right) The cathode assembly on the gravitational-tensioning holder after removing the mandrel.}
\label{fig:cathode-install1}
\end{figure}

When wrapping the foil, one side of the foil was initially aligned using the holes on the mandrel to ensure straightness and to make sure the conductive side of the foil was outside. After that, the foil was smoothed out on the mandrel to eliminate any remaining wrinkles. The overlap region was then glued. For the next step, the inner ring was glued to the foil, as can be seen in Fig.~\ref{fig:cathode-install1}. The seam of the foil was then aligned with the one inlet gas port in the inner ring. The inner ring contains four gas inlets to flow the drift gas inside the detector. Then, two copper tabs were glued into the DownStream end using conductive epoxy to establish electrical connections to the cathode foil. Finally, the outer ring was glued on the cathode foil. The assembly was then transferred to a holder assembly using an actuator, and held in a vertical position. Using gravity, the cathode surface was kept with minimal wrinkles as shown in Fig.~\ref{fig:cathode-install1}.

\begin{figure}[h!]
\centering
\includegraphics [width=0.23\textwidth,height=0.45\textwidth] {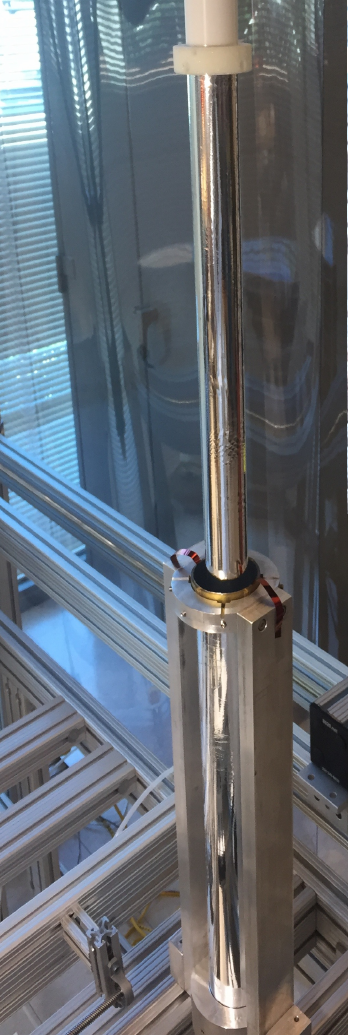}
\includegraphics [width=0.23\textwidth,height=0.45\textwidth] {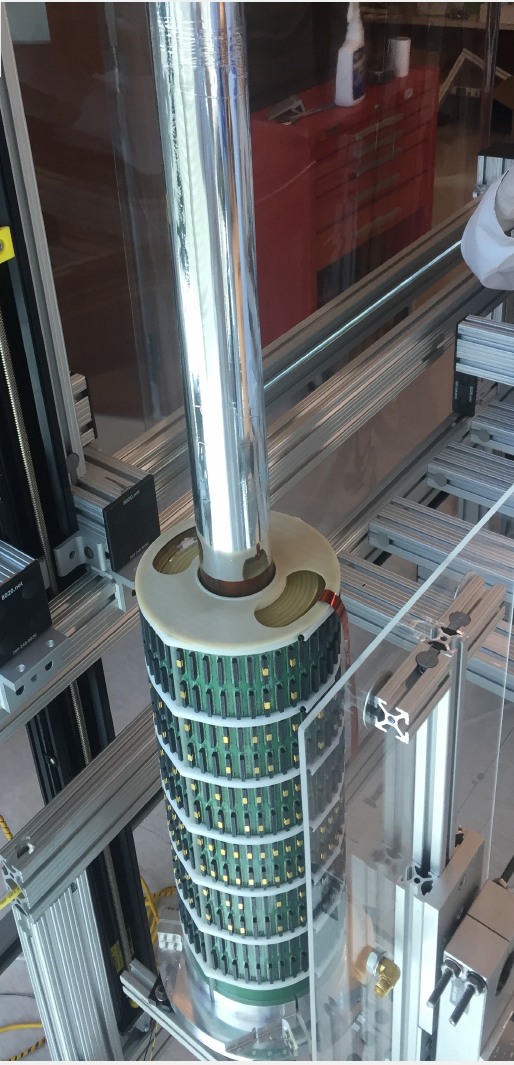}
\caption{(Left) Picture of the ground foil on the mandrel being lowered into the cathode assembly. (Right) The completed cathode assembly being lowered into the RTPC.}
\label{fig:cathode-install}
\end{figure}

The ground foil was sized a little longer along the cylinder axis using the same wrapping method as for the cathode foil. A foam ring was glued on the DownStream end of one side of the ground foil. This foam ring has gas inlets that connect to the cathode outer ring. Then, the mandrel with the tensioned ground foil was attached vertically to the actuator, and slowly inserted into the cathode holder assembly, as can be seen in Fig.~\ref{fig:cathode-install}. The foil was inserted a little more into the cathode assembly and the glue was applied on both ends. Thereafter, the mandrel was pulled back until the foam ring aligned with the edge of the cathode ring. Then, the entire cathode/ground assembly was lifted from the holder using the vertical actuator. At this point, any excess foil on the gas inlets was removed and a copper tape glued on the outer ring to make electrical contact. Finally, the cathode assembly was inserted into the detector in consecutive small steps as shown in Fig.~\ref{fig:cathode-install}.  

\subsection{\label{sec:TargetCons} Construction and Integration of the Target} 

\begin{figure}
\centering
    \includegraphics[scale=0.15]{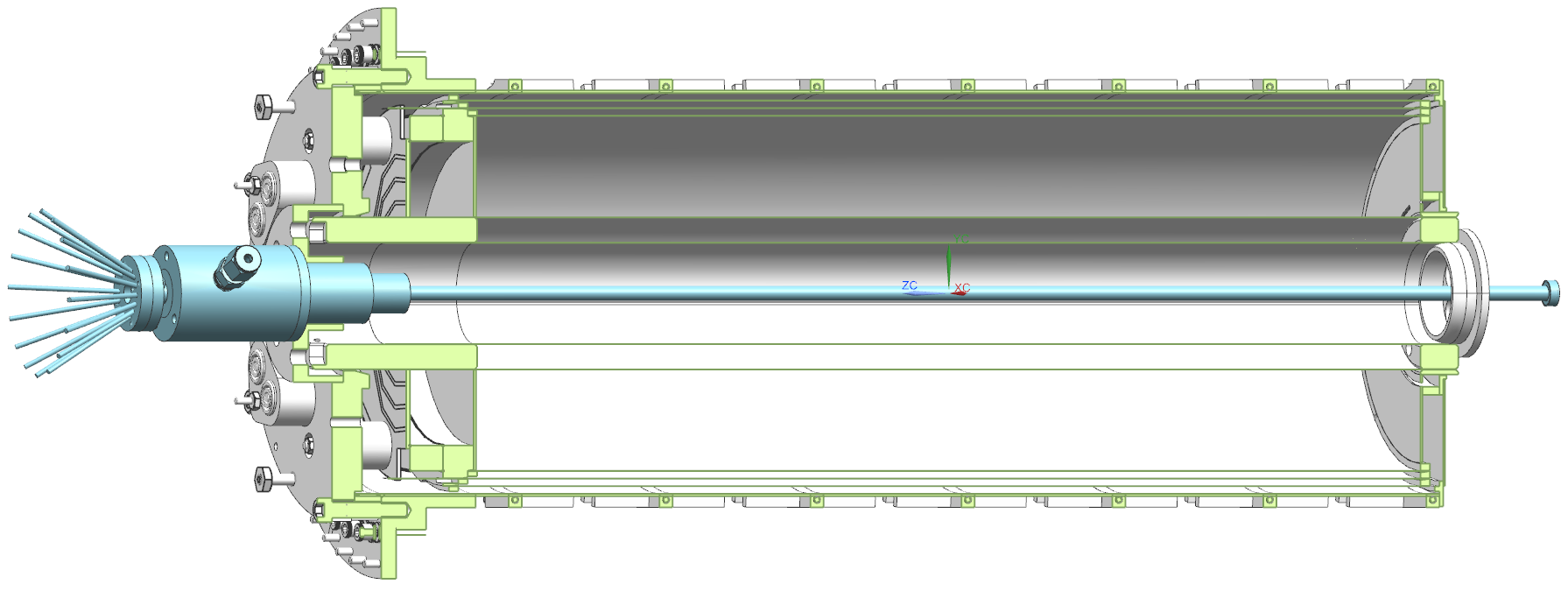} 
\caption{Schematic cross sectional drawing of the BOnuS12 RTPC showing the target system along the longitudinal axis of the detector.}
\label{fig:target_cad}
\end{figure}

The 6~mm diameter target straw was fixed along the center axis of the RTPC as shown in Fig.~\ref{fig:target_cad}. The assembly consists of a specially designed gas inlet tube, holder brackets, and the target straw itself. The assembly was installed inside the RTPC detector at the final stage, i.e., after the HV and leakage test of the detector. The upstream side of the target was fixed inside the RTPC by a specifically designed holder bracket and the downstream side was centered by using two styrofoam semi-disks with a center bore for the target.

\section{\label{sec:clas12} Integration into CLAS12}
\begin{figure*}[t!]
\centering
\includegraphics[scale=1.8]{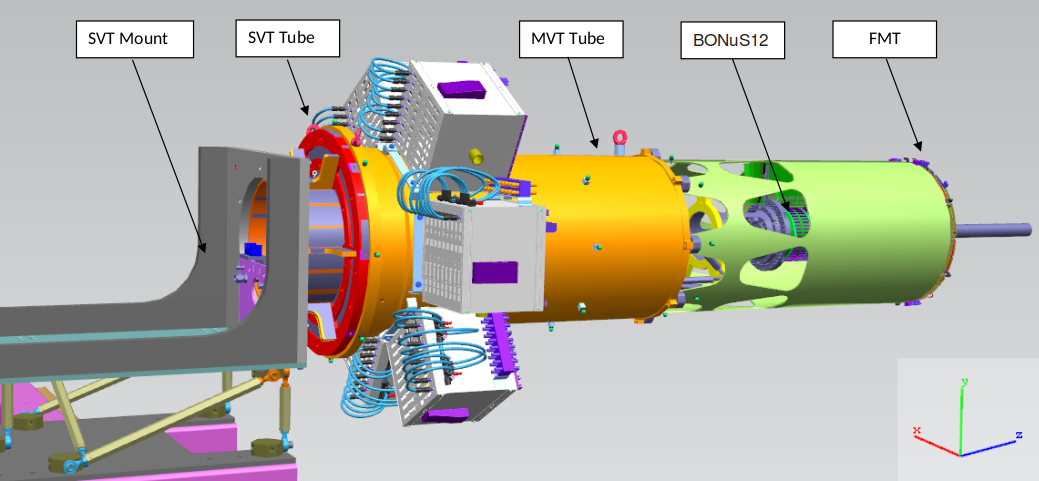}
\caption{CAD drawing for the assembly of the BONuS12 RTPC with the Forward Micromegas (FMT) on the main CLAS12-CVT cart.} 
\label{fig:bonus_clas12_cart}
\end{figure*}

The BONuS12 installation process was split into two stages. In the first stage, the BONuS12 RTPC, in addition to an upgraded Forward Micromegas Tracker (FMT), was installed on the so called CLAS12-CVT cart in a clean room outside Hall B (see Fig.~\ref{fig:bonus_clas12_cart}). In the second stage, the detector assembly was moved to  Hall B to be installed inside the CLAS12 Central Detector. 

\subsection{\label{sec:mechanics} Assembly and Installation Inside CLAS12} 
\begin{figure}[h!]
    \centering
    \includegraphics[scale=0.06]{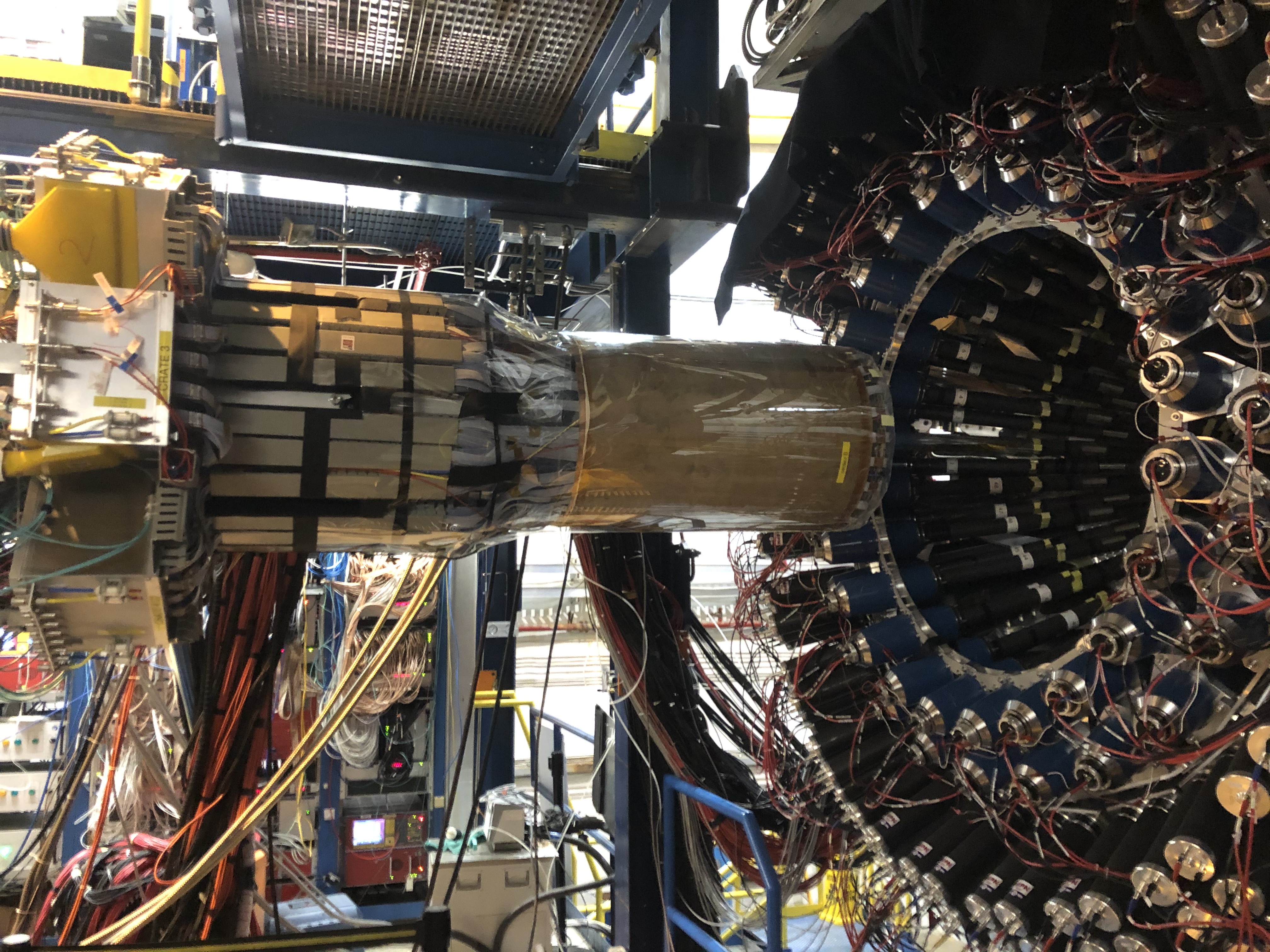}
    \caption{Picture showing the final assembly of the BONuS12 RTPC on the cart before being inserted in its ultimate position inside the central CLAS12 detector.}
    \label{fig:bonus12_installation}
\end{figure}

In order to perform the alignment of the BONuS12 RTPC, the BONuS12 target, and the upgraded FMT, the installation process followed these steps, see Fig.~\ref{fig:bonus_clas12_cart}:

\begin{itemize}
    \item After testing the functionality of the RTPC with cosmic rays, the target straw was installed inside the detector. 
    \item The RTPC was installed on the Micro-Vertex Tracker (MVT) tube.
    \item The RTPC was aligned with respect to the (MVT) tube and then RTPC and target system were surveyed.
    \item The RTPC was cabled and tested with cosmic rays. 
    \item The FMT detector was installed, aligned, cabled, and finally tested with cosmic rays.  
    \item A field shaper was installed around the whole RTPC-FMT assembly and wrapped with a plastic container to prevent any leaking helium from reaching the Photo-Multiplier Tubes (PMTs) of the Central Detector. The whole cart assembly was then moved to the experimental Hall B.
    \item The whole detectors-cart assembly was installed inside the core of the CLAS12 Central Detector. Fig.~\ref{fig:bonus12_installation} shows a picture taken at this stage in Hall B. 
    \item HV cables were connected to the BONuS12 RTPC and the FMT, and gas connections established. The detectors were tested for functionality using cosmic rays. 
    \item One final survey was performed at this stage to position the center of the BONuS12 RTPC at the center of the CLAS12 solenoid and to locate the whole assembly with respect to the CLAS12 spectrometer.
\end{itemize}

\subsection{\label{sec:gas} Gas Panel and Ancillary Devices} 
\begin{figure*}[hbt!]
    \centering
  \includegraphics[width=\textwidth]{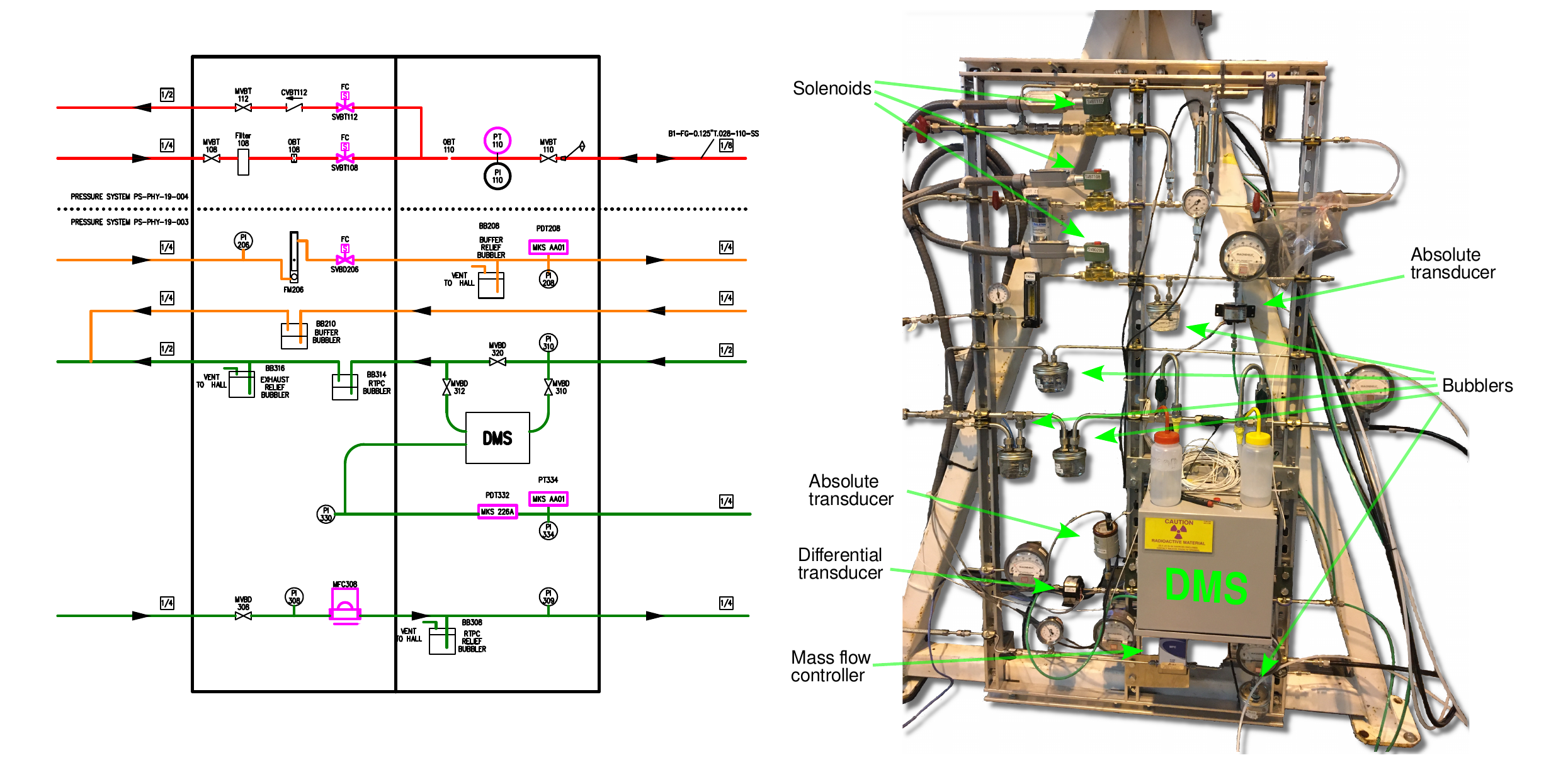}   
    \caption{(Left) CAD drawing of the BONuS12 gas panel. The green lines indicate the  drift region pre-mixed gas lines. The orange lines indicate the Helium-4 lines for the buffer region. The red lines indicate the gas lines for the target system. (Right) An image of the gas panel for the drift and buffer region, showing the active components and the safety passive parts (bubblers).}
    \label{fig:gas_panel_bonus12}
\end{figure*}

A custom gas panel was designed to supply a pre-mixed gas to the drift region of the RTPC and to a Drift Monitoring System (DMS), $^{4}$He gas to the buffer volume, and various gases to the target system. Fig.~\ref{fig:gas_panel_bonus12} shows a CAD drawing in addition to a photo of the gas panel. 

During the experiment, the RTPC drift region gas was supplied from high pressure, 2000~psi, pre-mixed cylinders of 20\% CO$_{2}$ in $^{4}$He by volume. Two cylinders were connected at all times in order to maintain the same gas composition during the experiment. Pressure regulators were used to reduce the gas supply pressure to 15 psi for the mass flow controller. Flow limiting orifices were used to limit gas flow in case of component failure. The same line-structure was used to feed the $^{4}$He gas to the buffer region inside the RTPC. In order to maintain the purity of the gases inside both the drift region and the buffer volume, the gases were flowing at rates to exchange one volume per hour. With this in mind, the flow rate of the drift region was about 150~sccm and about 20~sccm for the buffer volume.

For the target system, four gas bottles were connected to the gas panel, H$_2$, D$_2$, N$_2$, and $^4$He. The target straw was connected to the gas system on one end, preventing gas change by flowing gas through the straw. Hence, the gas was changed by venting the target straw to reach atmospheric pressure and then refilling the straw up to 5.8 atmospheric pressure. When changing the target gas to a different type, the target was emptied and refilled three times with the new gas to remove the previous gas. A flammable gas regulator was installed on the D$_2$ and H$_2$ bottles, while a non-flammable gas regulator was installed on the N$_2$ and $^4$He bottles. In each case a hose was connected to the supply gas panel. Both regulators were set at 86~psig. A 0.050~inch orifice was installed upstream of the regulators to limit gas flow during regulator failure. 100~psig relief valves were installed on the gas panel, before the solenoid valves. A flashback arrestor was installed after the flammable gas regulator to prevent flames from reaching the flammable gas bottles during an unexpected event. Additionally, a 0.010~inch diameter orifice was installed at the entrance of the target gas panel to reduce the flow during system flushing, and a second 0.001~inch diameter orifice before the straw to reduce the flow into and out of the straw and so minimizing stress on the straw and windows. Finally, a vent line from the target gas panel was set up to discharge to the outside of the Hall. 
In order to monitor both the drift and the target gas systems remotely, read-back values from the flow controllers, pressure, and temperature sensors were added to a slow controls system, and alarms were set for pressure and temperature ranges in the experiment control system. Pressure and temperature data for both target and drift gas systems were saved for offline studies.

\subsection{\label{sec:cables} Adapter Boards and Signal Transfer Cables} 

\begin{figure}[h!]
    \centering
    \includegraphics[scale=0.06]{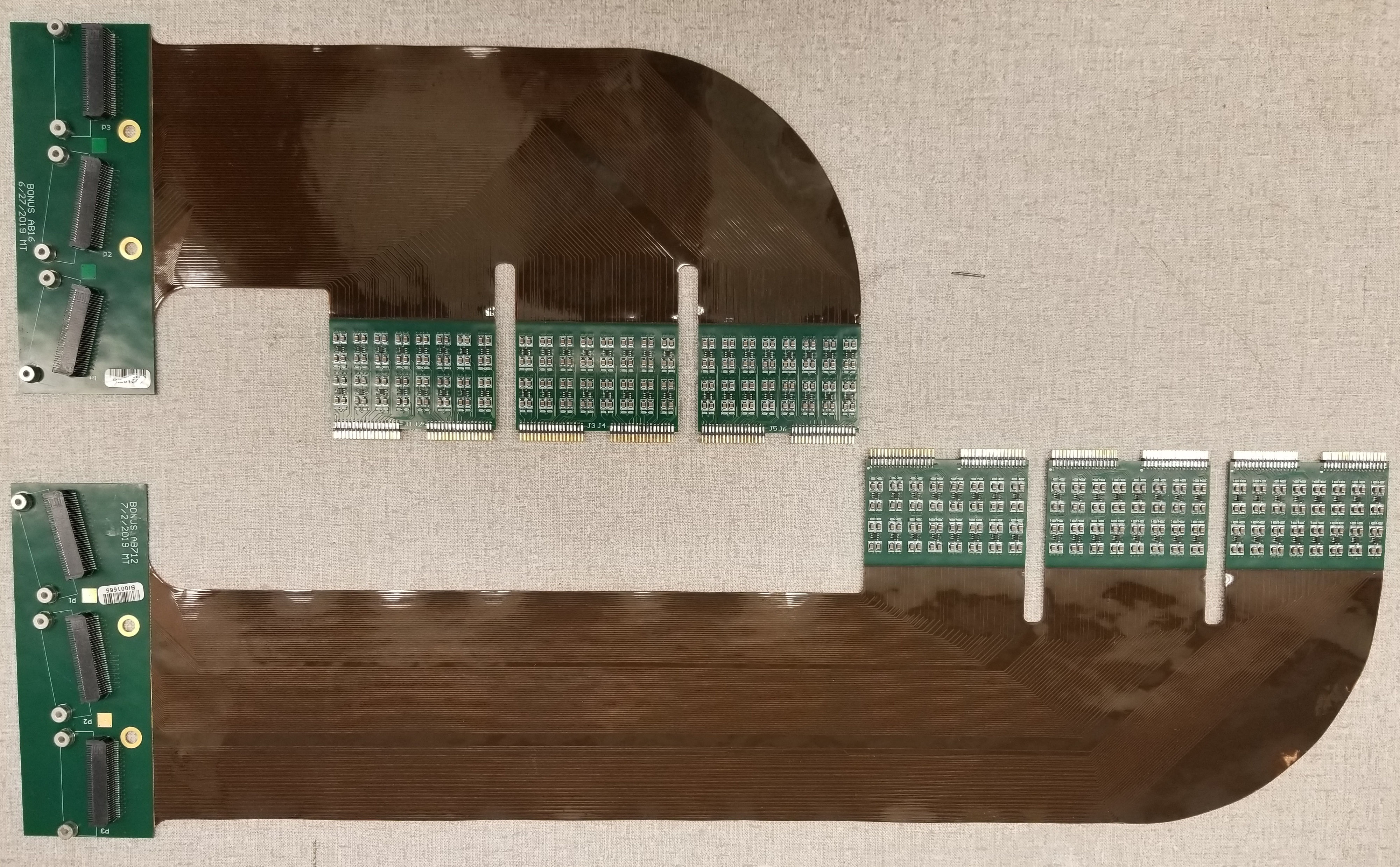}
    \caption{Two differently sized adapter boards to cover a complete row of connectors in the detector.}
    \label{fig:adapter_board}
\end{figure}

Signals from the RTPC readout pads were sent to the DAQ electronics through two differently sized adapter boards, shown in Fig.~\ref{fig:adapter_board}. They were connected on one side to the readout padboard and on the other side to Hitachi micro-coaxial cables. The adapter boards protected the readout electronics from over currents that can occur in case of a spark in the detector. Each adapter board transfers signals from 192 readout pads. The Hitachi micro-coaxial cables (manufacturer reference code KZ12-262) transfer the signals from the adapter board to the Front-End Units (FEUs). Each ribbon-like assembly had 64 flexible micro-coaxial signal cables with mini edge cards terminating the cable assembly on each end. In this configuration, three Hitachi cables, of length $\sim1.5$~m, were used to transfer signals from each adapter board to an FEU. A total of 270 cables was used to fully read out the RTPC.  

\subsection{\label{sec:FEU} Front-End Electronics Units (FEUs)} 
 Each FEU is composed of eight Samtec MEC8 input connectors, eight DREAM (Dead-timeless Readout Electronics ASIC for Micromegas) chips ~\cite{DREAMmanual}, and an 8-channel flash ADC. Each DREAM chip hosts 64 input channels with each channel having an integrated Charge Sensitive Amplifier (CSA), a shaper/filter, a 512-cell Switched Capacitor Array (SCA) analog circular buffer, and a discriminator for trigger building. The gain of the amplifier is chosen by selecting a range of input capacitances among four possible values between 50~fF to 600~fF. A value of 600~fF was chosen for our detector. Similarly, the peaking time of the shaper can be set to sixteen different values between 50~ns to 900~ns. Event sizes of up to 255 samples per trigger with a sample size of 40~ns can be continuously sampled in a 512-cell switched capacitor array circular buffer, which acts as the event memory pipeline. In this configuration, the DREAM chip performs in a dead-timeless readout mode of up to 20~MHz for a trigger rate of up to 20~kHz~\cite{DREAMmanual}. Upon receiving the trigger, a programmable number of samples of all channels, corresponding in time to the event, is read out serially through a differential analog buffer. The sampling is not stopped during the readout process, which allows nearly dead-timeless operation.

The analog samples from the eight DREAM chips are digitized by an 8-channel 12-bit flash ADC (AD9222). The eight serial streams of digital data are delivered to a Field Programmable Gate Array (FPGA) hosted in the FEU board. The FPGA reads corresponding data samples from the DREAM chips and follows the data processing steps: first, the pedestals are equalized after serial-to-parallel conversion; second, for each sample, the coherent noise affecting the DREAM chip inputs is estimated and subtracted on a per-chip basis, and last, the signal is zero suppressed on a per-channel basis. The FPGA is also responsible for forming an event from the retained channel data and sending it to a Back-End Unit (BEU). FEUs are powered by a 5~V source. Each FEU consumes $\sim20$~W of power when all eight DREAMs operate together. The FEUs can operate in the residual magnetic field of up to 1.5~T of the solenoid in Hall B without any noticeable change of their functionality~\cite{MMattie,CLAS12mm}. A continuous flow of air is necessary to cool the FEUs at room temperature within the crates.

\subsection{\label{sec:DAQ} Data Flow and Cosmic Test} 

\begin{figure}[h!]
    \centering
    \includegraphics[scale=0.2]{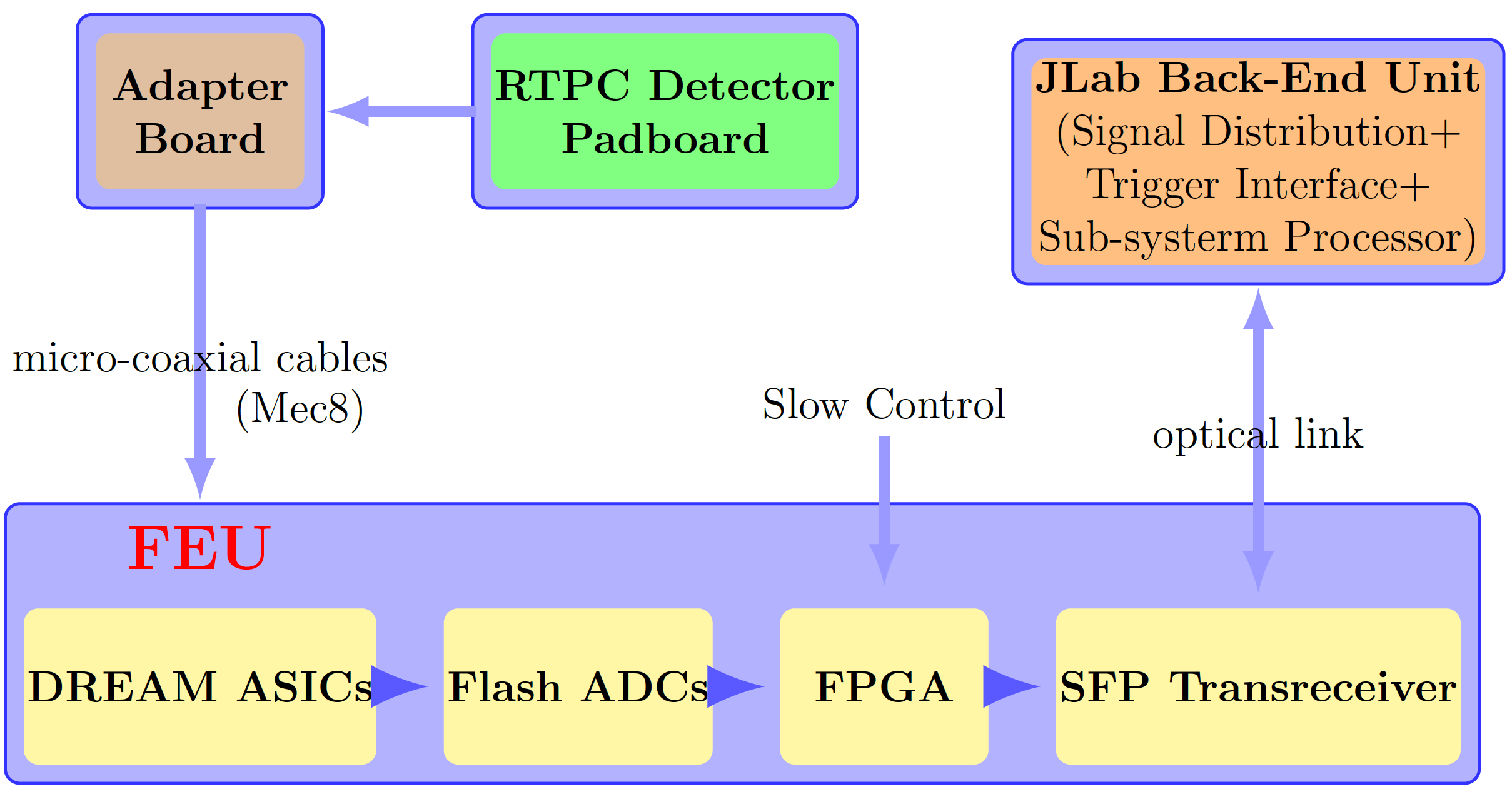}
    \caption{Data acquisition electronics and the flowchart~\cite{Jiwan2022}}
    \label{fig:daq_flowchart}
\end{figure}

\begin{figure}[tp!]
    \centering
    \includegraphics[scale=0.35]{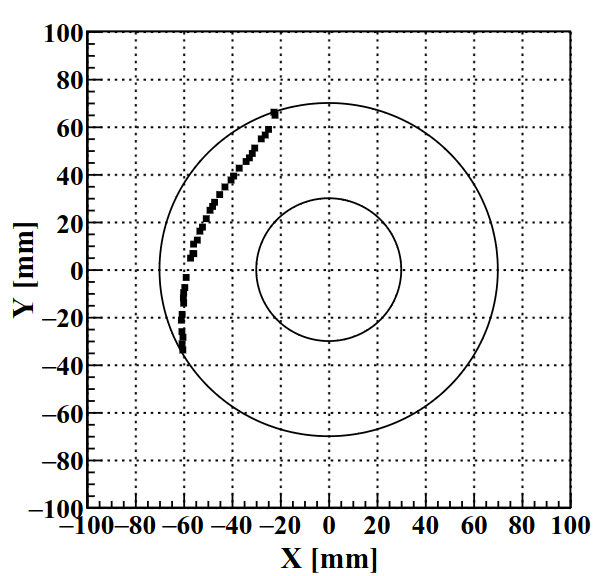}
    \caption{The Y vs. X cross sectional view for a detected cosmic ray track in the RTPC at 1.7~T solenoidal magnetic field.  }
    \label{fig:cosmic_track}
\end{figure}

During the RTPC prototype testing using cosmic rays, a few FEUs were connected directly to a computer via ethernet. In the final configuration, dedicated back-end units from the CLAS12 DAQ system were implemented to process FEU data and transfer them to the storage disks as illustrated in Fig.~\ref{fig:daq_flowchart}. The BEU was responsible for checking the data integrity, disentangling multi-event buffers, forming RTPC events corresponding to the triggers, and sending them to the CLAS12 Event Builder computer over a 10~GB/s ethernet link~\cite{CLAS12Daq}. Cosmic ray triggers were generated using a scintillator paddle and fed directly to the ECL trigger input pins on the front panel of the BEU electronics. Cosmic rays passing through the RTPC could be tracked with this setup and Fig.~\ref{fig:cosmic_track} shows an example of a cosmic ray detected in a cross sectional plane view of the RTPC at 1.7 T magnetic field..

\subsection{\label{sec:CODA} Integration into CLAS12 DAQ} 
The CLAS12 DAQ system was used during the experiment. A data readout trigger was generated by an electron being detected in the Forward Electromagnetic Calorimeter. The trigger information was then sent to the back-end units using optical fiber links~\cite{CLAS12Daq, CLAS12Trigger}. The RTPC used the fiber readout mode of two dedicated BEUs. Both BEUs distributed the global system clock, trigger, and synchronous commands to 36 FEUs, needed to read out all RTPC channels. For each trigger electron, the BEU performed local event building tasks by gathering data packets from all FEUs belonging to the same event (matching time stamps and event IDs). Data transfer rates of 200 MBytes/s were routinely achieved.

\subsection{\label{sec:Recon} Decoding and Reconstruction Software} 

The analysis of tracks recorded by the RTPC proceeds in the following steps:

\begin{enumerate}
    \item \underline{Signal Translation:} The data recorded from the RTPC consists of a sequence of readout pad ID and time slice information followed by the ADC readout, stored in an EVIO file. The decoder software translates the hardware information contained in the pad ID and time slice into a physical location on the pad and physical time relative to the trigger time. Together with the ADC information, this information is written to a file in HIPO format that contains all detector information for each triggered event.
    
   \item \underline{Track Finding:} The first task for the track finder is to sort all individual hits (consisting of pad ID, time stamp, and ADC value) into track candidates in each event. For this purpose, all hits are first sorted by time, and then hits within the same time slice are sorted together into ``proto-tracks'' whenever the corresponding pads are geometrically close. Hits within a time slice that cannot be added to an existing ``proto-track'' become the seeds of new ``proto-tracks''. Hits in subsequent time slices are added to an existing ``proto-track'' if it contains a hit that is a close neighbor in space and in time. The maximum distance in $\phi$ (8~degrees or 4 pad widths), $z$ (19~mm or nearly 5 pad lengths), and time (360~ns or 3 DREAM chip time slices) for a hit to be considered a close neighbor to another hit. These constants are defined and read from the CLAS12 software database. Hits that can be sorted to two different existing ``proto-tracks'' are assigned both track IDs, and eventually all track IDs that are shared by a common assigned hit are joined into hit-level track candidates. Given that the pad occupancy overall during a typical event is only a few percent within a time slice, most physical tracks are properly recovered with this procedure. However, tracks that cross close to each other in time and space end up being falsely combined and have to be disentangled.
   
   \item \underline{Disentangler:}
   As a first step, all hits on a single pad that have been sorted into one of the track candidates during the previous step are converted into single track points. This is done by calculating the ADC-weighted average time from all time slices with hits on that pad that belong to the given track candidate, and their overall ADC sum. After this procedure, each track candidate contains a list of unique pad positions linked to their average time and integrated ADC. As a next step, track candidates that might contain crossing physical tracks are flagged; this is based on several criteria:
   \begin{itemize}
       \item Overall length in time of the track (largest minus smallest average time) exceeds the maximum drift time.
       \item The time interval over which a pad saw a signal above threshold is longer than possible for real particle tracks.
       \item The track candidate contains points that are close in time but far apart in $\phi$ or $z$. 
   \end{itemize}
   Once a track candidate has been flagged, it will be passed to the disentangler. The disentangler sorts all points in the track candidate by time (from the latest to the earliest, corresponding to a track moving from the cathode towards the outer edge of the drift region). For each subsequent point on the track candidate, the disentangler uses the separation from the previous point in $z$ and $\phi$ as well as time to determine whether it really belongs to the same track. Additionally, a linear extrapolation of the existing track up to that point is used to predict the position of the next point, and points that don't match that position within predetermined limits (or that fail the previous test) are removed from the track and used as the start of a new track. At the end, each track candidate has been separated into one or more tracks that are likely belonging to different particles. However, real tracks might be falsely separated by this procedure, either if they contain too large gaps (perhaps due to inefficient pads) or if they curve back towards the cathode (``back-benders''). The latter case occurs when a particle has small enough momentum to miss reaching the edge of the drift region (GEM1) and instead is bent back towards the cathode by the magnetic field. Since the RTPC was built expressly to push the acceptance for protons to the lowest possible momenta, these tracks should be recovered. For this purpose, the disentangler is followed by a recombiner that checks whether the ends of each pair of individual new tracks match both in time and space, as well as in direction, and recombines them into a single track.
   
   \item \underline{Track Reconstruction:} The last step required to properly reconstruct a track in
   three dimensions is to relate the time of each point to its radial distance from the axis, and to correct each measured $\phi$-position for the Lorentz displacement of the drift electrons. Since not all tracks are in time with the trigger electron, the time offset between the track particle and the trigger time has to be determined. This is based on the observation that all tracks have to cross the cathode, and hence the latest time of a point on the track must correspond to the maximum drift time. After calibrating the detector (see Sec.~\ref{sec:drift}), this maximum drift time is known and can be used to calculate the offset in time, $t_{diff}$. The measured time of each point is corrected for this time offset, and the actual drift time corresponding to that point is converted into the distance from the outer edge of the drift region (GEM1). Extensive simulations using Garfield++ (see Sec.~\ref{ssec:Garf} below) and studies of actual reconstructed tracks have shown that, in spite of the large magnetic field along the axis of the detector, the {\em radial} part of the drift velocity is very near proportional to the radial electric field. Since the electric field between two concentric cylindrical conductors falls off like $1/r$, where $r$ is the radial distance from the detector/target center, one gets:
   \begin{equation}
       \frac{dr}{dt} = c\frac{1}{r} \Rightarrow
       \Delta(r^2) = 2c \Delta t.
   \end{equation}
   Solving this equation with the boundary conditions $r^2(t=0) = r_{max}$ (zero drift time from the outer edge of the drift region to the first GEM) and $r^2(t=t_{max}) = r^2_{min}$, one arrives at the simple formula:
   \begin{equation}
       r=\sqrt{r_{max}^2-(r_{max}^2-r_{min}^2)\frac{t}{t_{max}}}.
   \label{ttor}
   \end{equation}
   Here, $r_{max}$ is the radius of the anode (GEM1) cylinder, $r_{min}$ is the radius of the cathode, and $t$ is the offset-corrected measured average hit time, while $t_{max}$ is the maximum drift time from cathode to anode. 
   
   Once the radial position of a point is known, the displacement in $\phi$ of the drift electrons due to the Lorentz angle in the magnetic field can be corrected for. To a good approximation, the velocity component in the azimuthal direction is proportional to the radial velocity component, multiplied by the constant magnetic field, since the overall drift velocity is proportional to the Lorentz force acting on the drifting electron. This results in a largely constant Lorentz angle $\theta_L$ between the radial direction and the actual direction of propagation: 
    \begin{equation}
       v_{\phi} = r \frac{d \phi}{dt} = \tan{\theta_L} \frac{dr}{dt} \Rightarrow
       d \phi = \tan{\theta_L} \frac{dr}{r}.
   \label{Lorentz1}\end{equation}
   Integration on both sides yields for the total drift in $\phi$:
   \begin{equation}
       \Delta \phi = \Delta \phi_0 + \tan{\theta_L}\ln{\frac{r_{max}}{r}},
   \label{Lorentz}\end{equation}
 where second term on the r.h.s. is the change in the azimuthal coordinate between the ionization point and the intersection of the drift path with the anode (GEM1).
   $\Delta \phi_0$ accounts for the additional sideways drift between the anode (GEM1) and the padboard. 
   Again, simulations and analysis of well-reconstructed real tracks have shown that this simple formula works very well - see Sec.~\ref{ssec:Garf} below.
   
   After applying these two conversions, each point on each track has been assigned a unique position in three-dimensional space; for the $z$-position, the center in $z$ of the individual pad, which recorded the hit, is used. Since the magnetic field is very close to parallel to $z$ over the entire volume of the RTPC, and the electric field is uniformly radial thanks to the field shapers on both ends, it is a reasonable assumption that all drift paths are at constant $z$. 

    \begin{figure*}[tp!]
    \centering
    \includegraphics[trim={0 10cm 0 0},clip,scale=0.75]{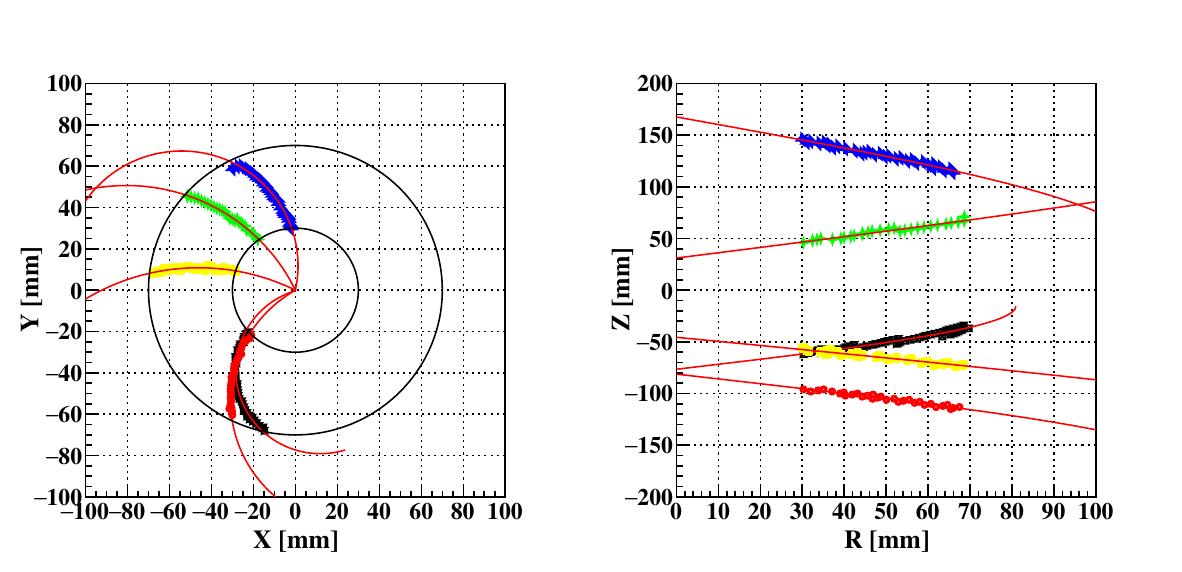}
    \caption{An example for reconstructed RTPC tracks in an experimental event. On the left: Y vs. X coordinate showing the hits of the tracks and the fits to the individual tracks (solid red line). On the right: The view for the reconstructed tracks in the Z vs. R plane, where Z is the position along the RTPC and R is the radial distance from the center of the target. }
    \label{fig:exp_tracks}
    \end{figure*} 
   
   \item \underline{Track Fitting:}
   Once a track has been reconstructed, the next step is to fit a particle trajectory to the set of hits that make up the track. The first-order approach is to assume that the particle trajectory is a helix, {\em i.e.}, a circle of radius $R$ in the $x-y$ plane overlaid with a constant velocity motion along the z-axis. This is the expected trajectory of a charged particle with fixed momentum in a constant magnetic field. The very slight variation of the magnetic field along the z-axis can be ignored as a good approximation, but the energy loss of the charged particle in the drift gas can be significant and is corrected for in a second step. The helix fitter used is a modified form of the one used already for the 6~GeV BONuS experiment \cite{PhysRevC.89.045206} and is based on a numerical recipe from ``Computer Physics Communications''~\cite{Helix}. It minimizes the reduced $\chi^2$ for the deviation of all hit positions from the fitted helix, with an additional constraint to minimize the distance of closest approach (DOCA) from the beamline. For tracks that do not intersect the anode and instead bend back towards the cathode (``back-benders''), only the first (outwards moving) part of the track is fit to determine the helix parameters, since the second part is likely affected significantly by energy loss and is not used for track reconstruction. Fig.~\ref{fig:exp_tracks} shows an example of the reconstructed RTPC tracks, from a production D$_2$ run using 10.4~GeV beam energy, along with the fits to these tracks. The tracks and their fits are shown in XY and RZ views of the RTPC. 

   The transverse (to $z$) part of the particle momentum can then be extracted as $p_\perp = qBR$, with the particle charge $q$, magnetic field $B$ and fit radius $R$. The fit also determines the slope 
   $dz/d\ell = \cot{\theta}$ and hence the scattering angle $\theta$ relative to the beam axis ($\ell$ is the path length along the track, projected onto the $x-y$ plane). From this, the magnitude of the particle momentum can be calculated as $p = p_\perp / \sin{\theta}$.
   
   The full length of the track between its minimum and maximum radial distance from the beam axis is $L=\sqrt{\Delta\ell^2 + \Delta z^2}$, where $\Delta\ell$ is the length of the track projected on the x-y plane, and $\Delta z$ is the distance along the $z$-axis between the first and last hit. The ADC values from all hits on the track are added and the sum is divided by the track length to get a quantity that is proportional to the average energy loss per unit length of the particle inside the drift region. (This assumes that the signal strength is proportional to the number of ionization electrons, which in turn is proportional to the energy loss).

\end{enumerate}

\begin{figure}[h!]
\centering
\includegraphics[scale=0.32]{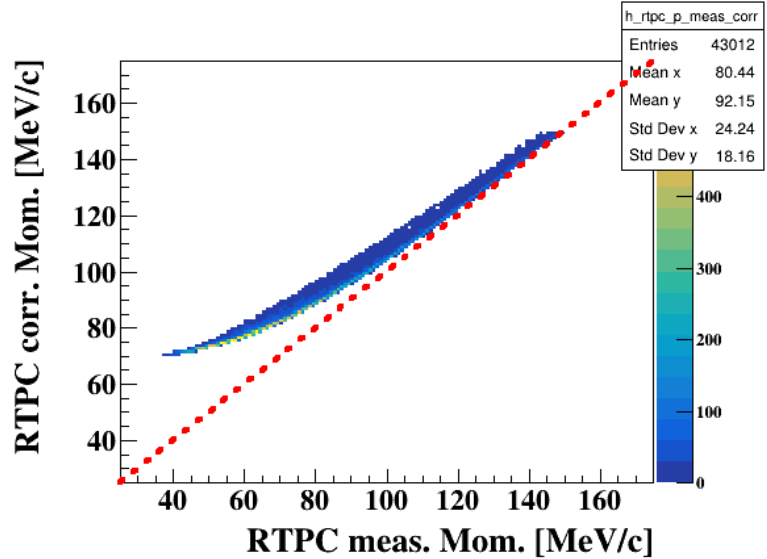}
\caption{The reconstructed corrected momentum of the RTPC detected protons versus the reconstructed measured momentum from the helix fit. The red-dashed represents the line of equality. }
\label{fig:rtpc_mom_corr}
\end{figure}

\begin{figure}[h!]
\centering
\includegraphics[scale=0.4]{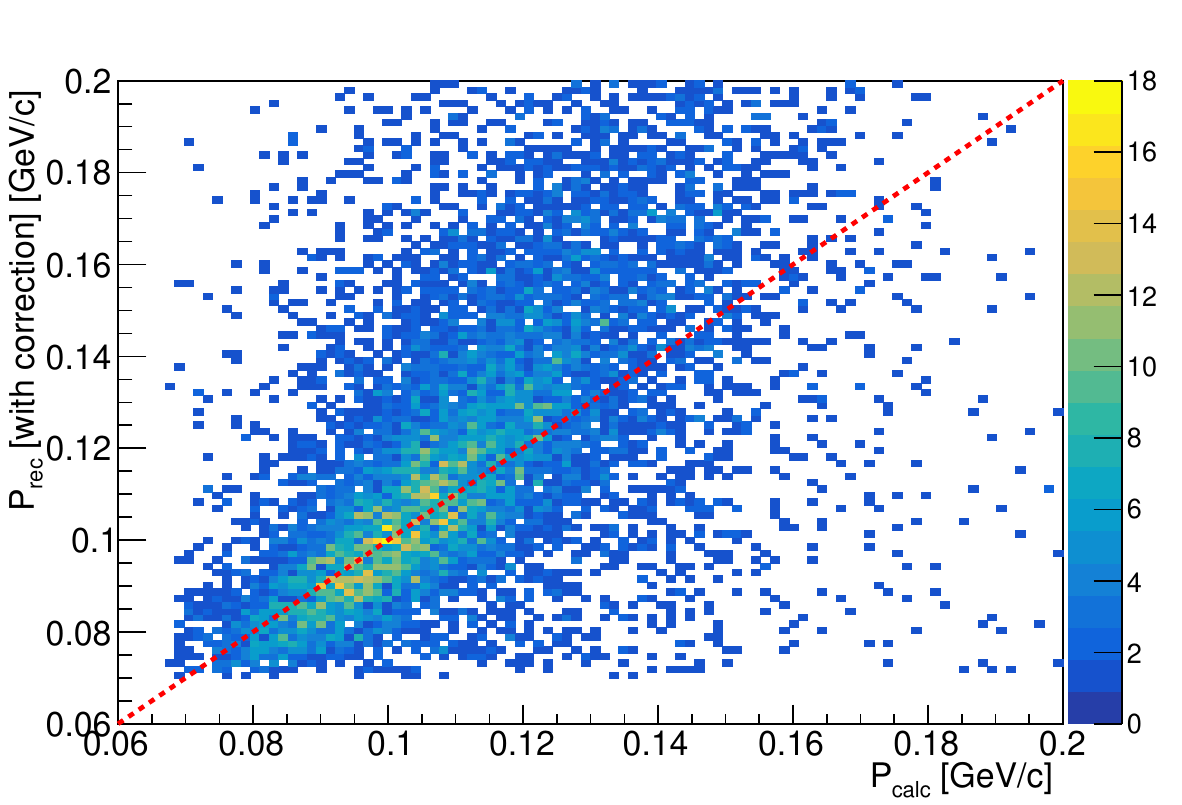}
\caption{ The correlation between the reconstructed corrected momentum of the RTPC protons versus the calculated  ``true'' momentum using the scattered electrons in the forward CLAS12 detector. The red-dashed line represents the line of equality.}
\label{fig:rtpc_meas_cal}
\end{figure}

   Finally, an empirical momentum correction was developed to account for the energy loss of the particle before entering the drift region. Most of this energy loss occurs close to the beamline, both inside the target gas and in particular inside the target walls. Therefore, one can use the reconstructed momentum from the helix fit as good approximation for the momentum after exiting the target, and can reconstruct the momentum at the vertex by applying the following parameterized correction to the momentum:
   \begin{equation}
    p_{corr} = p_{rec} + \frac{a_0} { \left(1.0 + \left( {\huge \frac{p_{rec}}{a_1} } \right)^2  \right)^ 3};\\
    \end{equation}   
with
\begin{equation*}
   a_0 = 56.45 - 0.007803 * \left( 1.0 + \left(\frac{|\theta_{rec} - 90.0|}{1.1765}\right)^{2}\right),\\
    \end{equation*}   
and
\begin{equation*}
    a_1 = 85.08 + 0.0003627 * \left( 1.0 + \left(\frac{|\theta_{rec} - 90.0|}{1.1045}\right)^{3}\right),
    \end{equation*}   
where $p_{rec}$ and $\theta_{rec}$ are the particle's reconstructed momentum, in MeV/c, and polar angle, in degrees, resulting from the helix fit, and $p_{corr}$ is the reconstructed corrected momentum of the particle at the vertex. Fig.~\ref{fig:rtpc_mom_corr} presents the correlation between the reconstructed corrected momentum of the recoil protons and the reconstructed measured momenta from the helix fit. Fig.~\ref{fig:rtpc_meas_cal} shows the correlation between the reconstructed corrected RTPC momentum of protons from elastic radiative $ep$ scattering as a function of their true momentum calculated using the kinematics of the detected final-state electron from the 2.1~GeV commissioning data on a H$_2$ target. 

\subsection{\label{sec:GEMC} Drift Simulation} 
As mentioned in Section~\ref{sec:Recon}, one can study the drift electron parameters and their dependence on environmental variables by simulating the drift of the electrons in the crossed electric and magnetic fields via the CERN Garfield++ software package ~\cite{VEENHOF1998726}. Eq.~\ref{driftTime} parameterizes the drift time as:
    \begin{equation} \label{driftTime}
    t = t_{max}\frac{r_{max}^{2} - r^{2}}{r_{max}^{2} - r_{min}^{2}}
    \end{equation}
which is a re-arranged form of Eq.~\ref{ttor}. In Eq.~\ref{driftTime}, the parameter $t_{max}$ can be expanded as a polynomial in $z$ to account for the variations in the electromagnetic fields along the central axis of the detector. These parameters are then used in the Geant4 Monte Carlo (GEMC) simulation of the experiment and in the event reconstruction routines as well.

\subsubsection{\label{ssec:Garf} Simulation of Ionization in the Drift Region}
The Garfield++ simulation is split into two steps, the drift region from the cathode to the first GEM layer, and the transfer region from the first GEM layer to the readout pads. The maximum drift time ($t_{max}$) and Lorentz angle ($\theta_{L}$) can be extracted from the drift region, whereas the offset correction of the drift time ($t_{offset}$) and the additional sideways shift ($\Delta\phi$) can be evaluated from the transfer region. The standard running conditions of the RTPC during the experiment were: 

\begin{itemize}
    \item -4335~V relative voltage between the cathode and the first GEM foil ($V$). 
    \item 3.78~T solenoidal magnetic field along the $z$-direction ($B$). 
    \item 80\%/20\%, by volume, as the ratio of the drift gas components ($^{4}$He/CO$_{2}$).  
    \item 753~Torr gas pressure in the drift region ($P$).  
    \item -300~V relative voltage in the transfer regions between the GEM foils and between the last GEM foil and the padboard.
    \item 20~°C as the running temperature ($T$).  
\end{itemize}

For the drift region, one can simulate the drift of ionization electrons released at different values of radial distance ($r$) from the center of the target (see Table.~\ref{tab:Table_1}), and for various $z$ positions along the center axis under standard running conditions. Samples of 1,000 events for each simulation iteration were simulated. The drift time and angle are the results of the mean values from these events in each simulation set. Table~\ref{tab:Table_1} shows the results for the drift time at the center of the target ($z$= 0~cm) for the different $r$ values. To obtain $t_{max}$, the data is fitted with Eq.~\ref{driftTime}, in which $t_{max}$ is a freely adjustable parameter in the fitting function, determined by finding the minimum $\chi^{2}$ for the data set. As a result, a $t_{max}$ of 2972~ns, at $z$= 0~cm, was obtained. Similarly, for the parameters of the drift angle, Eq.~\ref{Lorentz} is used as the fitting function, in which $\tan \theta_{L}$ is the free parameter. At $z$= 0~cm, $\tan \theta_{L}$ was measured to be 0.9206~radian.

\begin{center}
\begin{table}
\centering
\begin{tabular}{cccc}
\Xhline{1pt}
           & \multicolumn{2}{c}{drift time} &               \\ 
r {[}cm{]} & $\mu${[}ns{]} & $\sigma$[n]    & fitting value{[}ns{]} \\ \hline
3.0        &               &                & 2972      \\ 
3.1        & 2930          & 18             & 2927      \\ 
3.5        & 2731          & 17             & 2730      \\ 
4.0        & 2452          & 17             & 2452      \\ 
4.5        & 2136          & 17             & 2136      \\ 
5.0        & 1781          & 16             & 1783      \\ 
5.5        & 1392          & 14             & 1393      \\ 
6.0        & 964           & 12             & 966       \\ 
6.5        & 501           & 8              & 501       \\ 
6.9        & 103           & 4              & 103       \\ \hline
           &               & $\chi^{2}$     & 0.011         \\ 
\Xhline{1pt}
\end{tabular}
\caption{The drift time of the simulated ionization electrons 
at $z$= 0~cm for different radial ($r$) positions and the fitting results.}\label{tab:Table_1}
\end{table}
\end{center}

\begin{figure}[h!]
\centering
\includegraphics[scale=0.42]{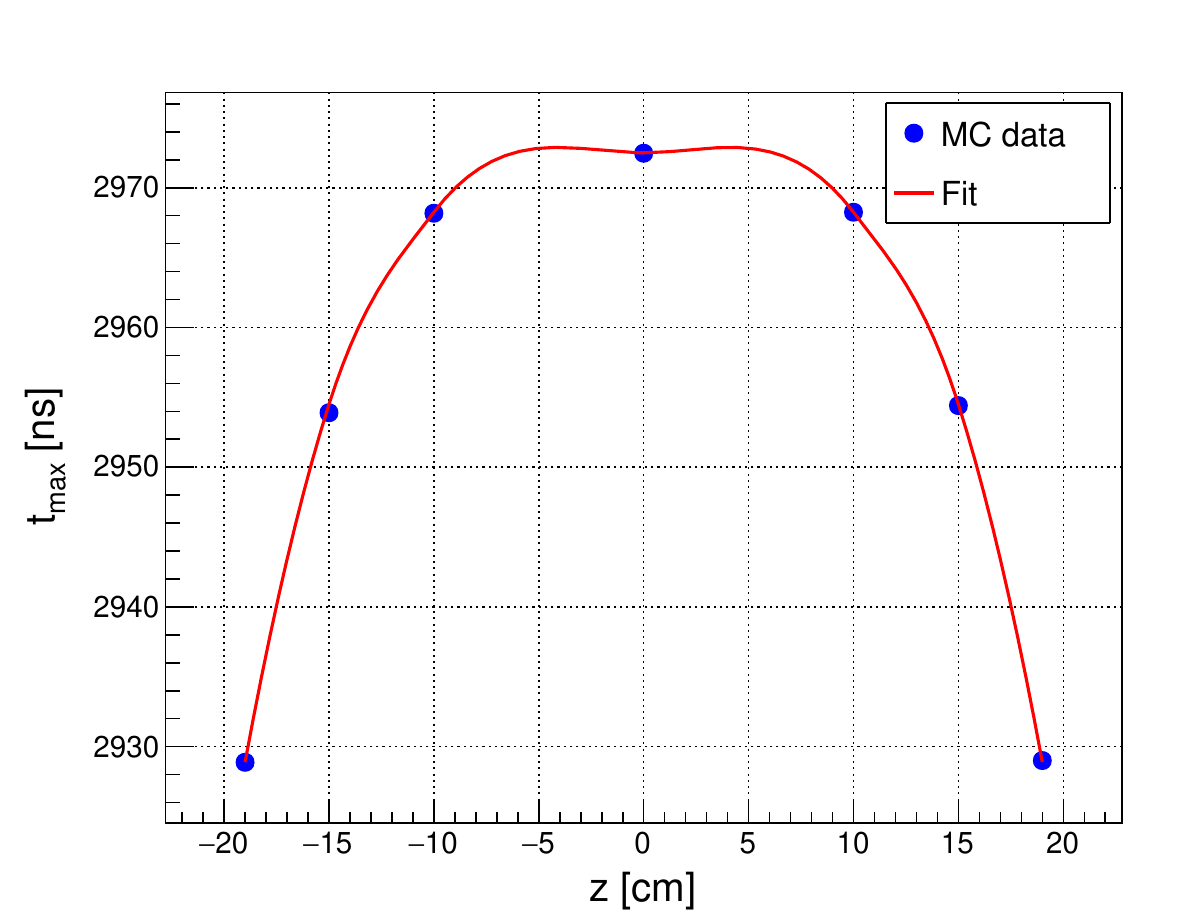}
\caption{$t_{max}$ vs. $z$-positions. The blue dots are the values extracted from the simulation and the red line is the fitting curve.}\label{fig:fig1}
\end{figure}

\begin{figure}[h!]
\centering
\includegraphics[scale=0.42]{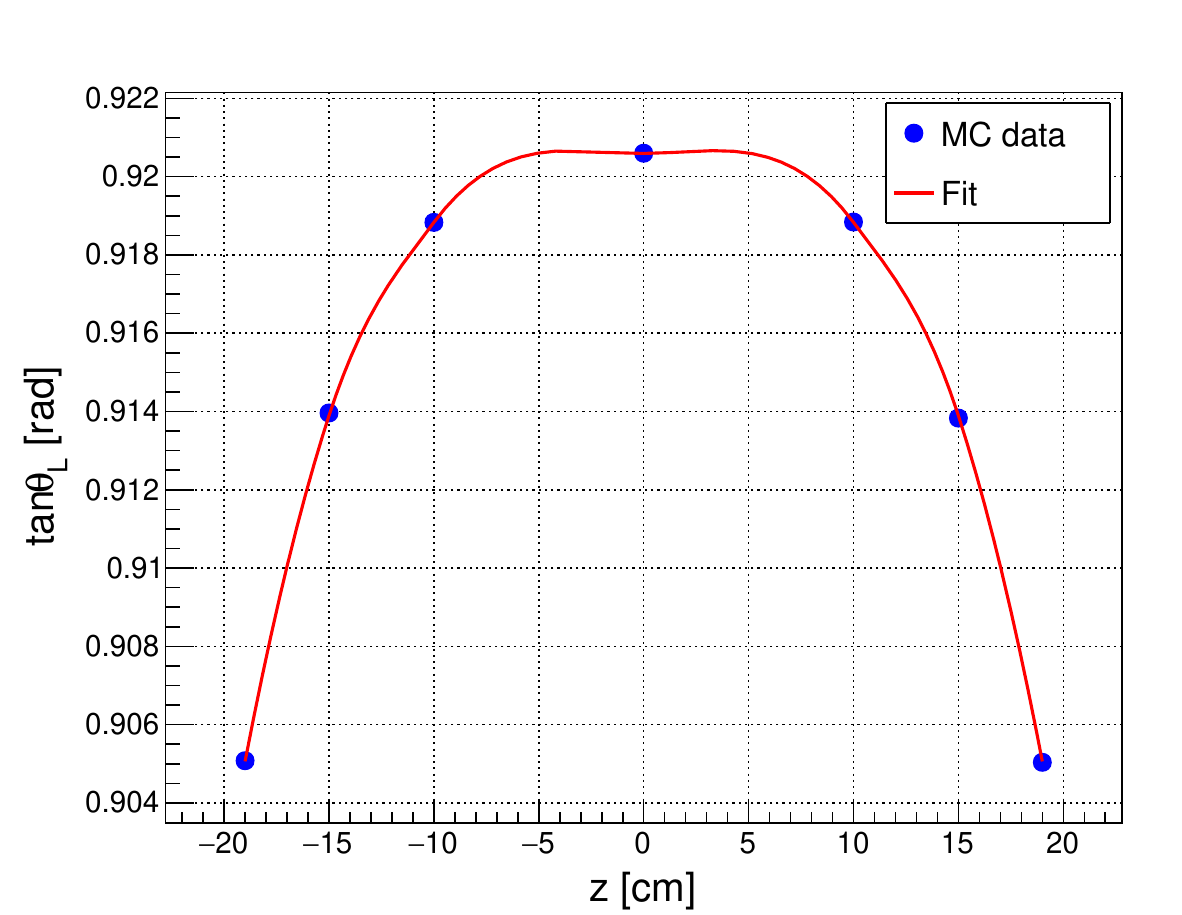}
\caption{$\tan \theta_{L}$ vs. $z$-positions. The blue dots are the values extracted from the simulation and the red line is the fitting curve.}\label{fig:fig2}
\end{figure}

Due to the finite size of the solenoid, the magnetic field might not be uniform along the $z$-direction. Therefore, the previous procedures were repeated at different $z$-positions: -19, -15, -10, 0, 10, 15, 19~cm. The fit parameters can be expanded as a second-order polynomial in $z^{2}$, $a_{t}(z) = A_t + B_t*z^{2} + C_t*z^{4}, a_{\phi}(z) = A_{\phi} + B_{\phi}*z^{2} + C_{\phi}*z^{4}$. Fig.~\ref{fig:fig1} and Fig.~\ref{fig:fig2} present the results for the drift time $t_{max}$ and the angle $\theta_L$, respectively. The data points are well described by this fit within the intrinsic uncertainties of the data points at each $z$ value, which is a fraction of a percent. 

\begin{figure}[h!]
\centering
\includegraphics[scale=0.42]{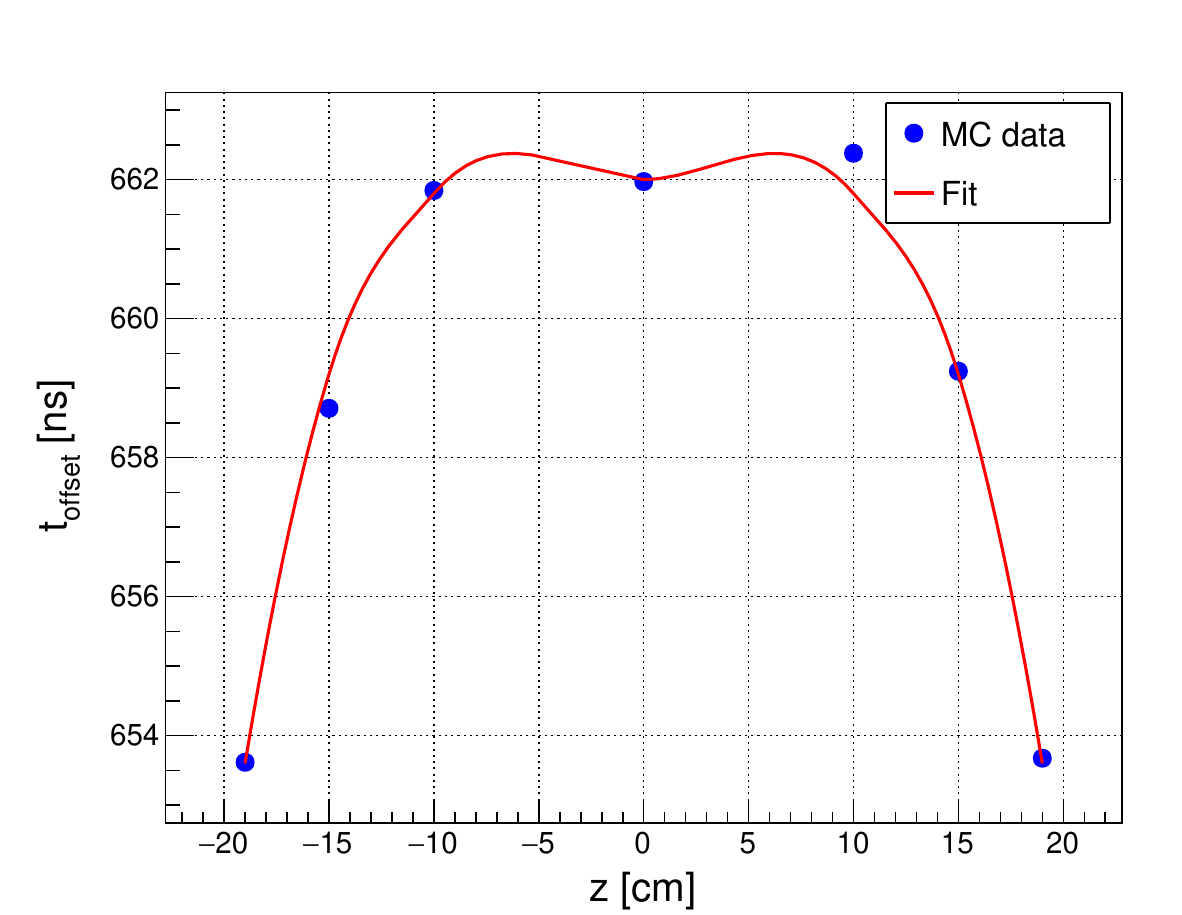}
\caption{The $t_{offset}$ vs. $z$-positions. The blue dots are the values extracted from the simulation and the red line is the fitting curve.}\label{fig:fig3}
\end{figure}

\begin{figure}[h!]
\centering
\includegraphics[scale=0.42]{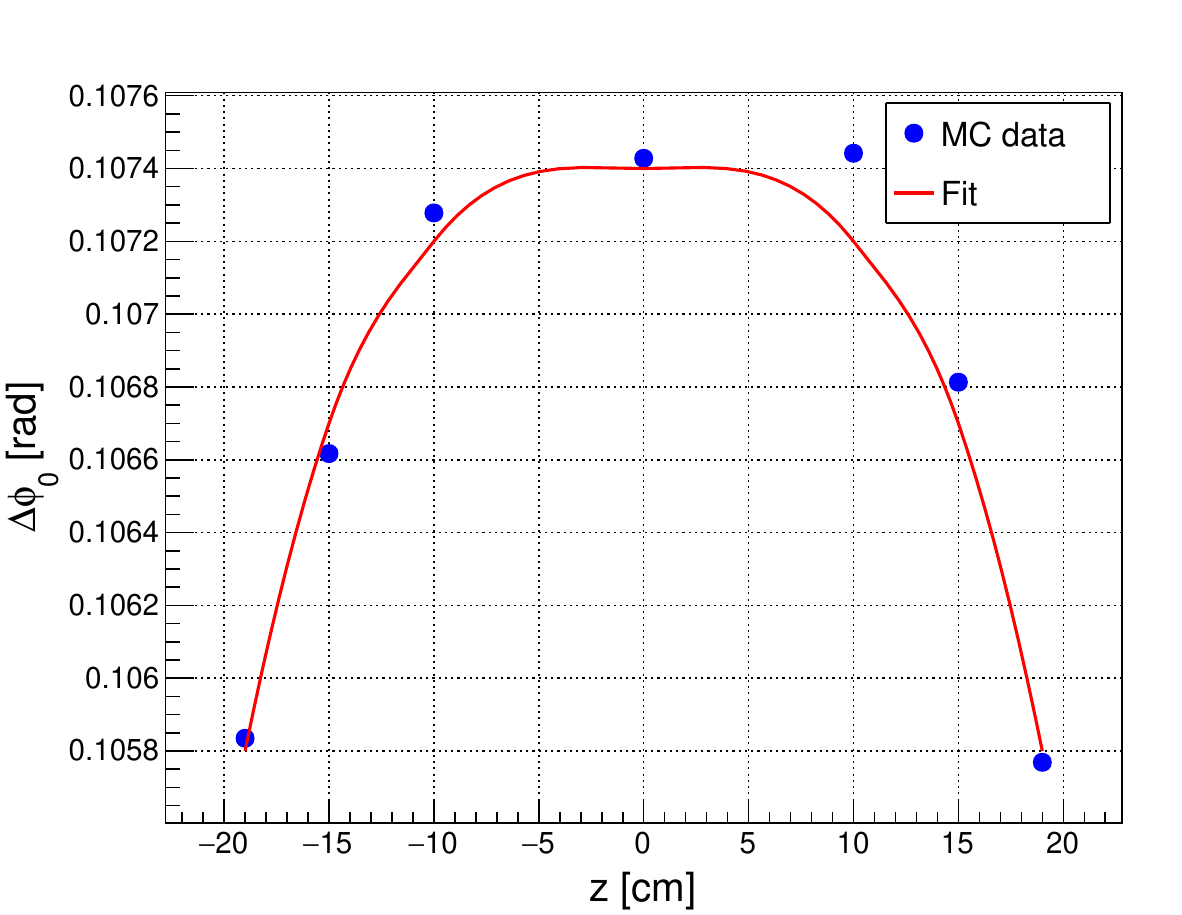}
\caption{The $\Delta\phi_{0}$ vs. $z$-positions. The blue dots are the values extracted from the simulation and the red line is the fitting curve.}\label{fig:fig4}
\end{figure}

For the transfer region, the time and the angle for the ionization electron drifting through the transfer region were simulated. There are two additional GEM layers dividing this transfer region into three sections, each separated 3~mm radially. Again, samples of 1,000 events for each section were simulated. Therefore, $t_{offset}$ and $\Delta\phi_{0}$ are the sum of the drift time and angle when the ionized electron passed through these three regions. Like for the drift region, the steps in the same z-positions are repeated and fitted these simulated data results with similar second-order polynomials in z² and the corresponding coefficients A, B, and C were extracted. Fig.~\ref{fig:fig3} and Fig.~\ref{fig:fig4} show the fit results for $t_{offset}$ and $\Delta\phi_{0}$, respectively.

Since the ionization electrons are sensitive to environmental conditions when drifting, simulations were carried out for the drift time and the drift angle at different cathode's voltage, $B$-field, gas mixture ratio, pressure, and temperature. For this part, our results were limited to $z$= 0 positions. Only one condition was changed in each simulation iteration, while keeping the others at the standard setting. The alternate conditions are: $V$= -4235~V (for transfer region, $V$= -330~V); $B$=~3.73~T; $^4$He/CO$_{2}$: 79\%/21\%; Pressure= 763~Torr; $T$= 25~°C. Similar to the standard conditions, the simulation results were fitted and the parameters from the fitting functions were extracted. These results are listed in Tables~\ref{tab:Table_7} to \ref{tab:Table_10}, showing the relative percentage change with respect to the standard conditions. These studies confirm that the changes to the drift time and drift velocity of the ionization electrons, due to non-perfect control over running conditions during the experiment, are within the expected reconstruction resolutions.

\begin{table}[h!]
\begin{tabular}{lcccc}
\Xhline{1pt}
Quantity &  Value change   & $t_{max}$ [ns]& Change [\%] \\ \hline
B field\,{[T]}    & 3.78 $\rightarrow$ 3.73  & 2936.34  & -1.22	       \\ \hline
Temp.\,{[°C]}     & 20 $\rightarrow$ 25      & 2968.80  & -0.12	       \\ \hline
Pres.\,{[Torr]}   & 753 $\rightarrow$ 763    & 2975.97  & 0.12	       \\ \hline
He\,[\%]          & 80 $\rightarrow$ 81      &  2967.82 & -0.16	       \\ \hline
HV\,{[V]}         & -4335 $\rightarrow$ -4235& 3041.88  &  2.34	       \\ 
\Xhline{1pt}
\end{tabular}
\caption{The $t_{max}$ sensitivity on the different experimental conditions. The $t_{max}$ relative change is calculated with respect to $t_{max}$ (=2972.47~ns) at the standard conditions. } \label{tab:Table_7}
\vspace{1.0cm}

\begin{tabular}{lcccc}
\Xhline{1pt}
Quantity       &  Value change  & $\tan\theta_{L}$ [rad]& Change [\%] \\ \hline
B field\,{[T]} & 3.78  $\rightarrow$ 3.73        & 0.9083        & -1.34     \\ \hline
Temp.\,{[°C]}  & 20 $\rightarrow$ 25             & 0.9349        & 1.55	     \\ \hline
Pres.\,{[Torr]}& 753 $\rightarrow$  763          & 0.9090  	     & -1.27	 \\ \hline
He\,[\%]       & 80 $\rightarrow$ 81             & 	0.9425       & 2.38	     \\ \hline
HV\,{[V]}      & -4335 $\rightarrow$ -4235       &  0.9217  	 & 0.12	     \\ \hline
\Xhline{1pt}
\end{tabular}
\caption{The $\tan\theta_{L}$ sensitivity on the different experimental conditions. The relative change is calculated with respect to $\tan\theta_{L}$ (=0.9206~rad) at the standard conditions.} \label{tab:Table_8}
\vspace{1.0cm}

\begin{tabular}{lcccc}
\Xhline{1pt}
Quantity       &  Value change & $t_\textrm{{offset}}$ [ns] & Change [\%] \\ \hline
B field\,{[T]} & 3.78 $\rightarrow$  3.73   &  654.74   & -1.10	    \\ \hline
Temp.\,{[°C]}  & 20 $\rightarrow$ 25        &  661.76   & -0.03	    \\ \hline
Pres.\,{[Torr]}& 753 $\rightarrow$  763     & 663.00    & 0.16	    \\ \hline
He\,[\%]       & 80 $\rightarrow$ 81        & 661.66    & -0.05	    \\ \hline
HV\,{[V]}      & -300 $\rightarrow$ -330    & 602.11    & -9.04	    \\ \hline
\Xhline{1pt}
\end{tabular}
\caption{The $t_\textrm{{offset}}$ in different variations and the relative change with respect to the standard conditions (= 661.97~ns).} \label{tab:Table_9}
\vspace{1.0cm}

\begin{tabular}{lcccc}
\Xhline{1pt}
 Quantity       &  Value change & $\Delta\phi_{0}$[rad.] & Change [\%]\\ \hline
B field\,{[T]}  & 3.78$\rightarrow$ 3.73    & 0.1060       & -1.32   	  \\ \hline
Temp.\,{[°C]}   & 20  $\rightarrow$ 25      & 0.1090	   & 1.48	      \\ \hline
Pres.\,{[Torr]} & 753 $\rightarrow$ 763     & 0.1059       & -1.38	      \\ \hline
He\,[\%]        & 80  $\rightarrow$ 81      & 0.1099       & 2.32 	      \\ \hline
HV\,{[V]}       &-300 $\rightarrow$ -330    & 0.1068       & -0.61	      \\ \hline
\Xhline{1pt}
\end{tabular}
\caption{The $\Delta\phi_{0}$ in different variations and the relative change with respect to the standard conditions (= 0.1074~rad).} \label{tab:Table_10}
\end{table}

\subsubsection{GEMC}

\begin{figure}[t]
\centering
\includegraphics[scale=0.5]{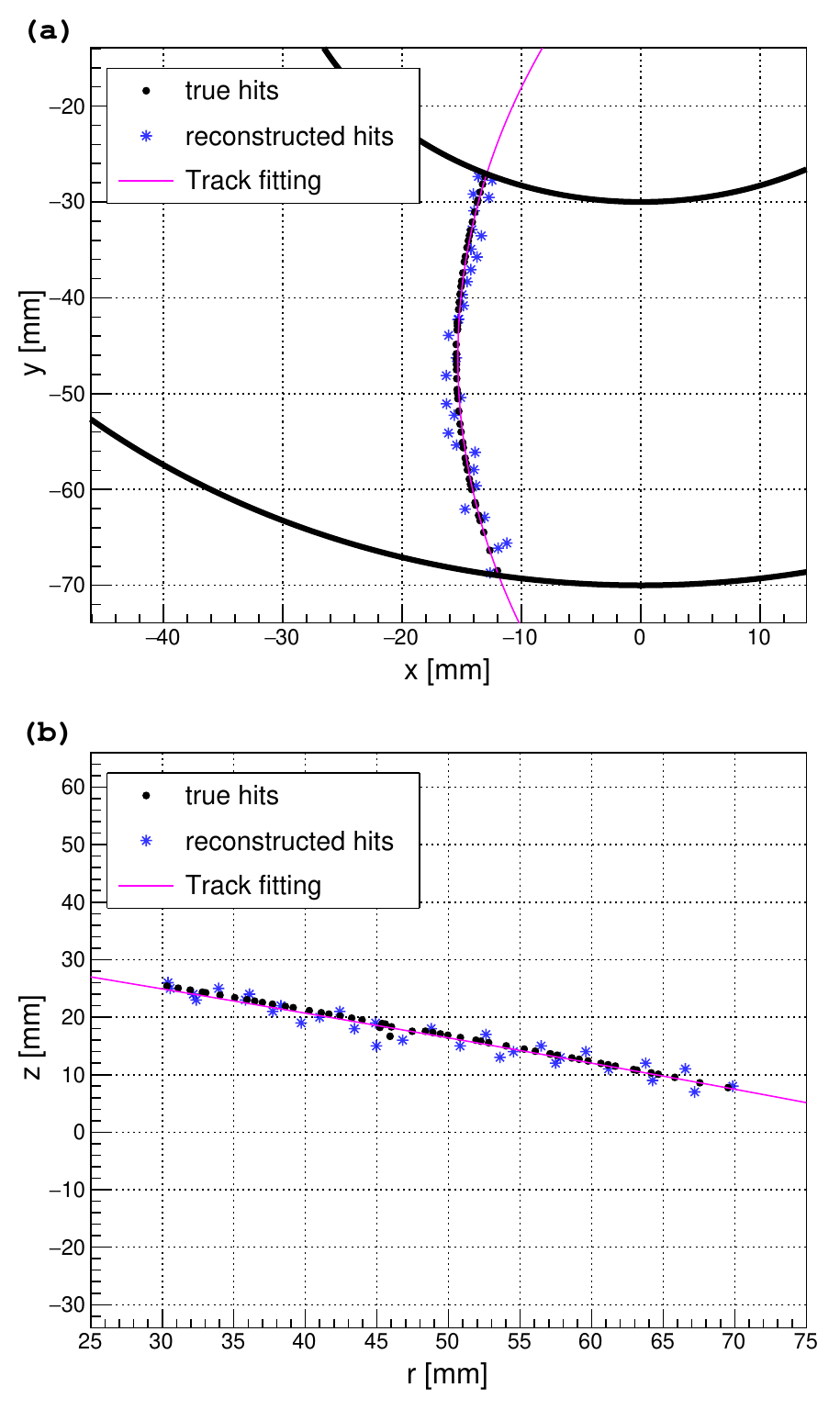}
\caption{Track comparison between the true hits simulated by GEMC (black dots) and the reconstructed hits (blue star) in (a) $x$-$y$ plane and (b) $r$-$z$ plane. The magenta line is a trajectory by track fitting based on the reconstructed hits.}
\label{fig:track_reconstructed}
\end{figure}

\begin{figure}[t]
\centering
\includegraphics[scale=0.5]{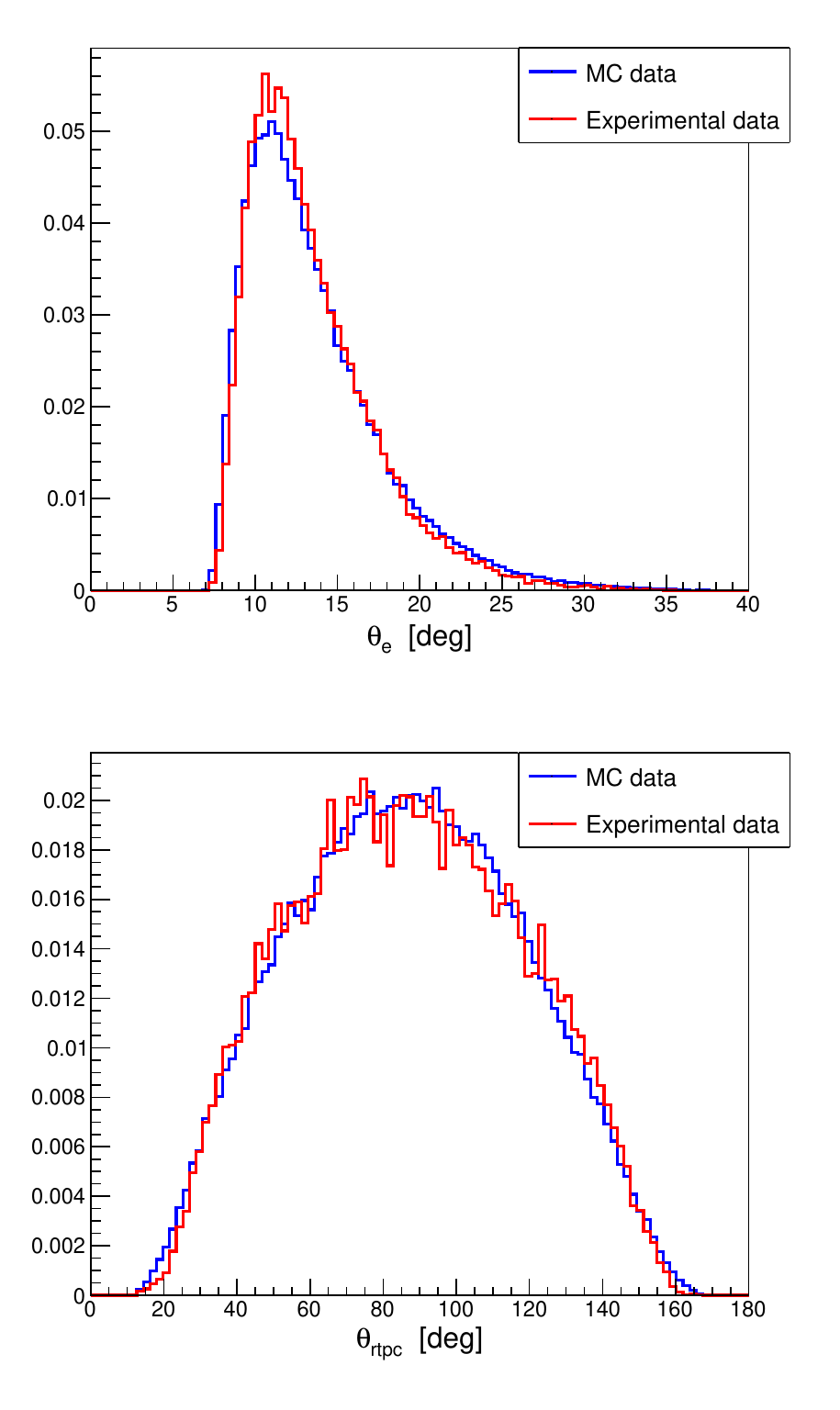}
\caption{Comparison between real data and Monte-Carlo simulated data in terms of the polar-angle distribution of the reconstructed electrons (top) and recoil-spectator protons (bottom) from neutron DIS events.}
\label{fig:comp_data_MC}
\end{figure}

To guide the data analysis and to simulate the passage of particles through the various CLAS12 detectors, a Geant4 Monte-Carlo (GEMC) package was used. The CLAS12 GEMC software package uses the Geant4 toolkit as the base to simulate the behaviour of particles passing through the material of the different detectors. See \cite{UNGARO2020163422} for details on the CLAS12 GEMC simulation package. By defining the geometry, material, and the physical layout of an experiment in GEMC, one can track particles through the interactions they make while crossing the different detectors. While Geant4 is fully sufficient to generate the ionization hits a recoil charged particle would make in the drift region of a gaseous detector, such as this RTPC, it does not simulate the drift of the low energy ionization electrons. For this reason, the Garfield++ results were used and fine-tuned using the experimental data, to parameterize the drift paths and the drift time of the ionization electrons inside the RTPC. The parameters were then implemented in the hit process in GEMC to evaluate how the ionization electrons drift to the readout padboard. Fig.~\ref{fig:track_reconstructed} compares a track simulated by GEMC and after its reconstruction. Fig.~\ref{fig:comp_data_MC} shows a comparison between real data and Monte-Carlo simulated data in terms of the polar-angle distributions of the electrons and the spectator protons from the reconstructed neutron DIS events.

\section{\label{sec:calibration} Calibration}
\subsection{\label{sec:drift} Drift Velocity and Drift Path Calibration}
As described in Section~\ref{sec:Recon}, the reconstruction routine requires information about the maximum drift time from the cathode to the anode ($t_{max}$ in Eq.~\ref{ttor}), as well as the time offset between the trigger time and the reconstructed time for a signal generated at the anode. The signal generating electrons drift through all three GEM layers before reaching the padboard, and then various cable delays, the internal signal shaping, and the clock synchronization between the trigger and the FEUs need to be considered. Furthermore, the Lorentz angle and the azimuthal offset which enter Eq.~\ref{Lorentz} are also required.

\begin{figure}[h!]
    \includegraphics[scale=0.45,trim=1mm 5mm 1mm 11mm ]{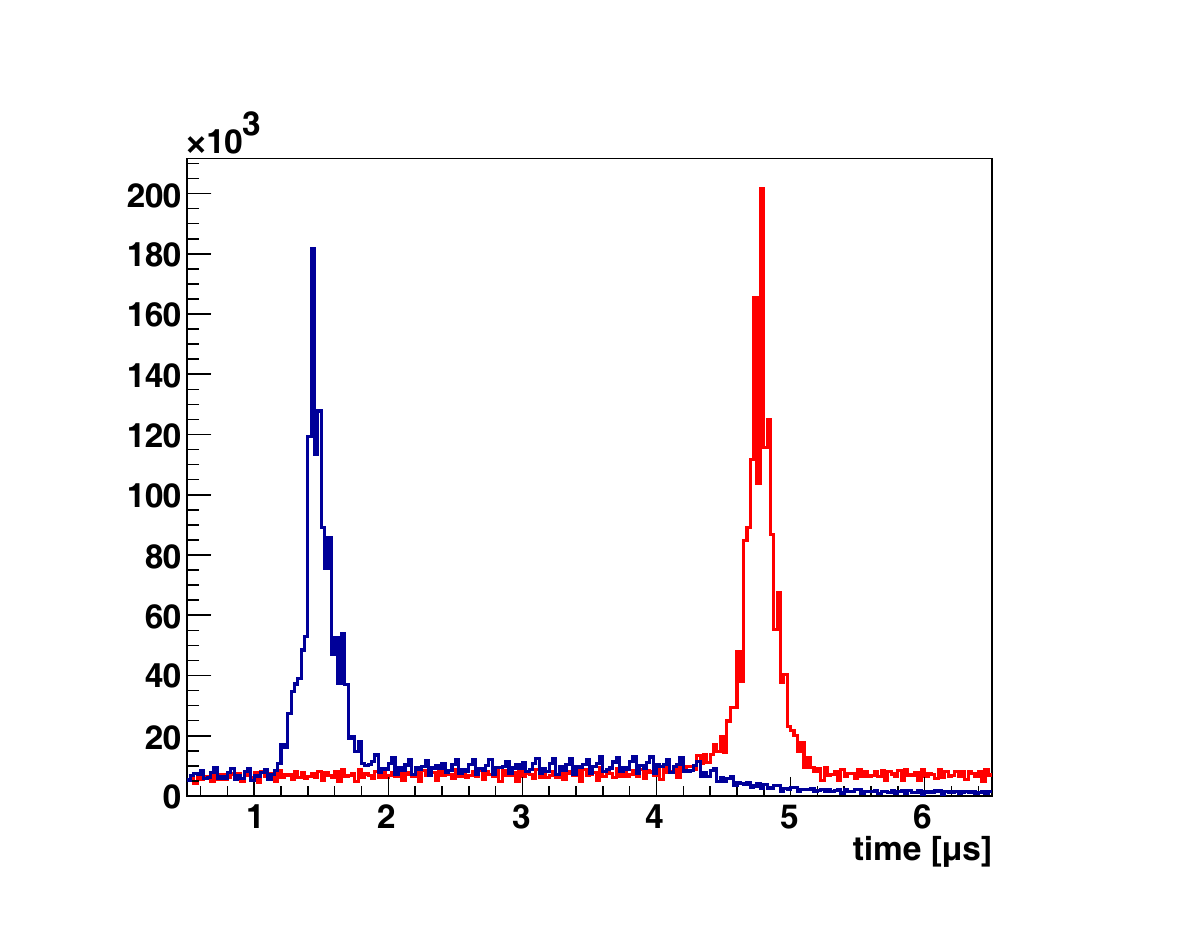}
    \caption{Distribution of the average time for the earliest (blue, left) and latest (red, right) hit on each reconstructed track, using 2.1~GeV commissioning run data on an H$_2$ target. The peak on the left corresponds to the time offset between the reconstructed time for the outermost hit (near the anode) of an in-time track (coincident with the trigger electron). The peak on the right is the average recorded time for the innermost hit (near the cathode) and corresponds to the offset plus the maximum drift time, $t_{max}$. The parameter $t_{max}$ is the difference in the centroids between these two peaks. The centroid of the left peak corresponds to the expected latest time for an in-time track; the track time offset $t_{diff}$ is determined as the difference between this centroid and the latest hit time measured for a given track.}
    \label{fig:timecalib}
\end{figure}

The time offset and $t_{max}$ are determined by monitoring the time of the earliest and latest hits in each track, see Fig.~\ref{fig:timecalib}. The corresponding distributions show two clear peaks due to in-time tracks that traverse the entire drift region from cathode to anode, in reasonable agreement with the values from the Garfield++ simulation. The centroids of these peaks are used for the relevant database parameters. 

The Lorentz angle and azimuthal offset are determined, to first order, from the Garfield++ simulation, Sec.~\ref{ssec:Garf}. Then, the radiative elastic events from the 2.1~GeV commissioning data on a pure hydrogen target are used to fine-tune these parameters by optimizing the agreement between the observed proton track and that predicted track from the observed electron.

\subsection{\label{sec:gain} Gain Calibration and Energy Loss Reconstruction} 
The main goal of the BoNuS12 experiment is to use the RTPC to identify spectator protons from the reaction $D(e,e'p_{s})X$. However, several kinds of particles make tracks with heavy ionization trails in the RTPC. The particle identification is performed using the measured $dE/dx$, the amount of energy the particle loses in the drift gas per unit of distance traversed. The RTPC is not sensitive, in production running, to particles with minimum ionization, because they would not make a dense enough ionization trail to give a signal above threshold. The proton has the smallest mass of all particles that could make a dense track, therefore particles with atomic numbers greater than one could  produce good tracks as well.

Experimentally, $dE/dx$ can be calculated from the collected ADC values as: 
\begin{equation}
 \left\langle \frac{dE}{dx} \right\rangle= \frac{\sum\limits_{i} \frac{ADC_{i}}{Gi}}{L},
\end{equation}
where the sum runs over all the pads contributing to a track. $ADC_{i}$ is the 
recorded amplitude in each pad $i$, and $G_{i}$ is its gain. The $L$ is the 
total visible track length in the drift region of the RTPC. The electron collection system of the RTPC has 17,280 readout pads. The gain of each pad is the ratio between the deposited energy and the output recorded value. The gains are extracted using two sequential steps:

   \begin{figure}
    \centering
    \includegraphics[scale=2.8]{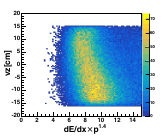} 
    \caption{A zoom on the proton's calculated $dE/dx$ (multiplied by the momentum ($p^{1.4}$) to linear the dependence on the momentum), using the set of gains extracted from the first method, showing the artificial dependence of $dE/dx$ on the position along the RTPC.}
    \label{fig:dEdx_1st}
    \end{figure} 

    \begin{figure*}[t!]
    \centering
    \includegraphics[scale=2.8]{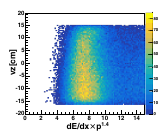}
    \includegraphics[scale=0.45]{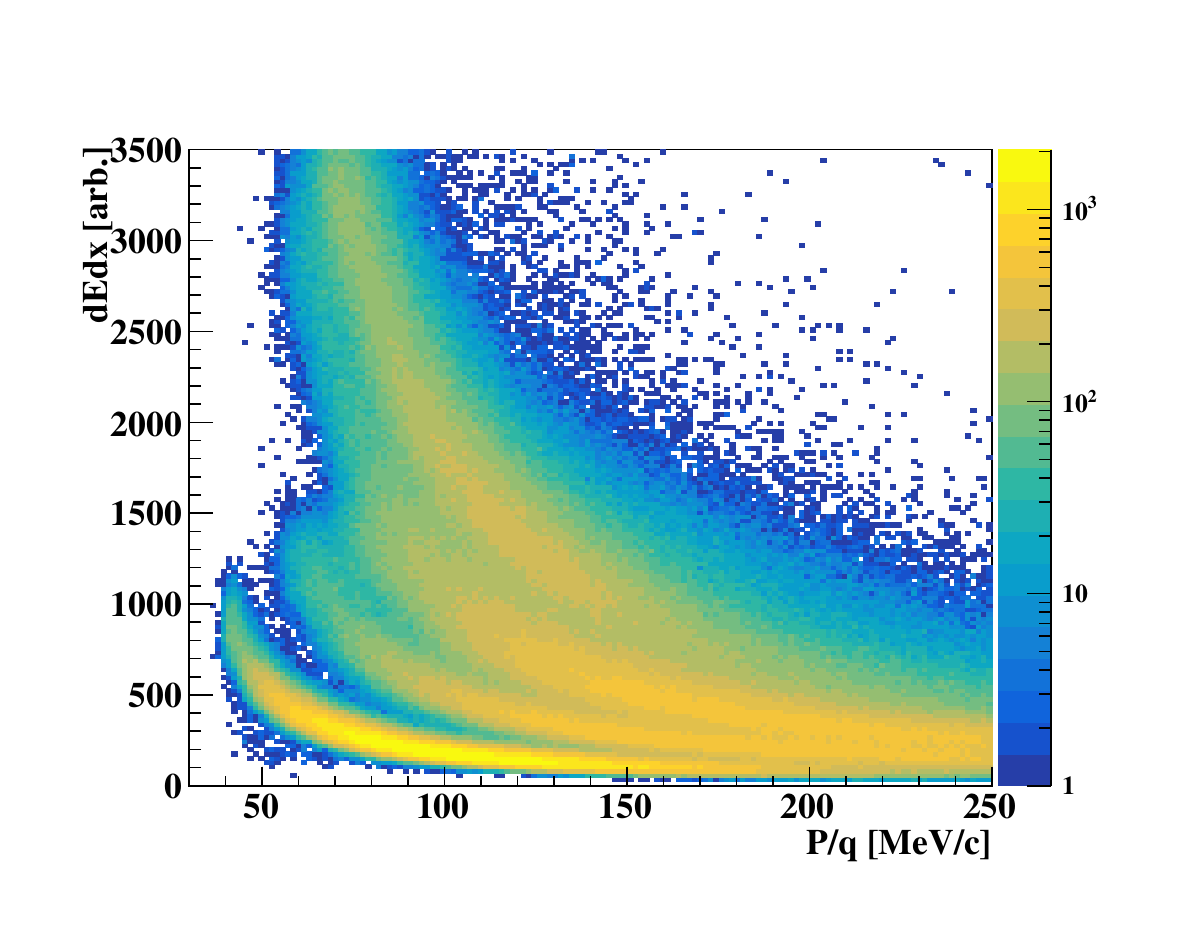}
    \caption{(Left) The relationship between $dE/dx$ (multiplied by the momentum ($p^{1.4}$) to approximately cancel the dependence on the momentum), calculated using the gains extracted from the second method, and the longitudinal position along the RTPC. (Right) The $dE/dx$ versus the measured momentum, per unit charge, for the different detected recoil: protons (lowest $dE/dx$), Deuterium, $^3$He/$^3$H, $^4$He (highest $dE/dx$).}
    \label{fig:dEdx_2nd}
    \end{figure*}

\begin{itemize}
    \item In the first step, the 
    ratio of the ADC sum for a given pad divided by the number of hits on that pad during an experimental run is used to define the gain $G_i$. While this method leads to a more homogeneous pad response, it did not correct for an observed dependence of $dE/dx$ of the recoils on the longitudinal position along the target. Fig.~\ref{fig:dEdx_1st} shows this artificial dependence of $dE/dx$ on the position along the detector.

    \item In the second step, the gain factors are further corrected using in-time reconstructed tracks spanning the full drift region from the cathode to the anode, by comparing every pad's collected charge (ADC value) on the track to the median collected charge along the entire track. Then, for every pad, the gain is defined as the average of this ratio over a large number of tracks. This second step did correct for the vz-dependence that was observed after the first step. Fig.~\ref{fig:dEdx_2nd} shows the new $dE/dx$ versus vz-dependence. Additionally, the right plot in this figure shows $dE/dx$ versus momentum per charge, for the different recoil particles from the smallest (proton) to the heaviest recoil ($^4$He).

\end{itemize}

\section{\label{sec:performance}Detector Performance}

\subsection{Resolution and Efficiency} 
\begin{figure*}[!ht]
    \centering
    \begin{overpic}[scale=0.4]{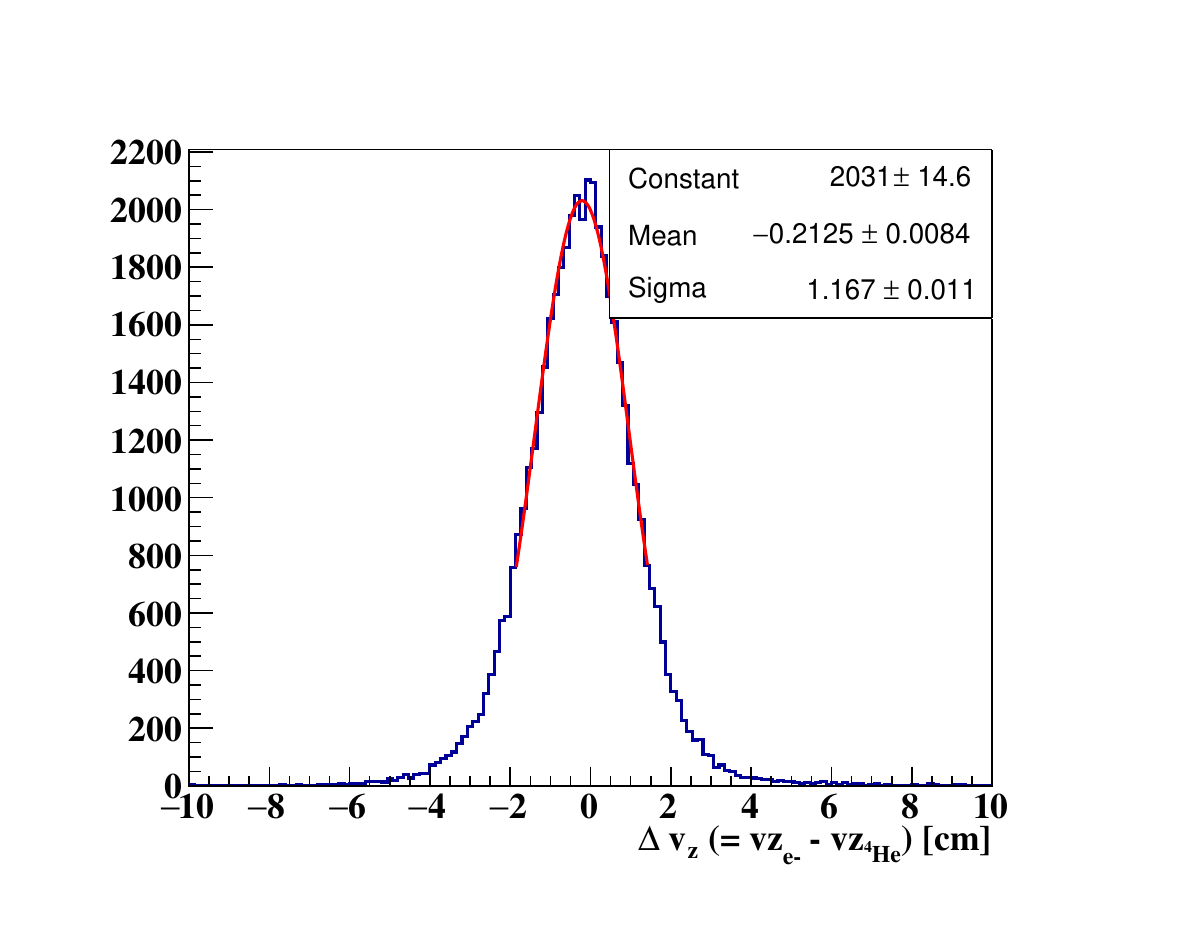}
    \put (20,50) {\Large(a)}
    \label{fig:resolutions_a}
    \end{overpic}
    \begin{overpic}[scale=0.4]{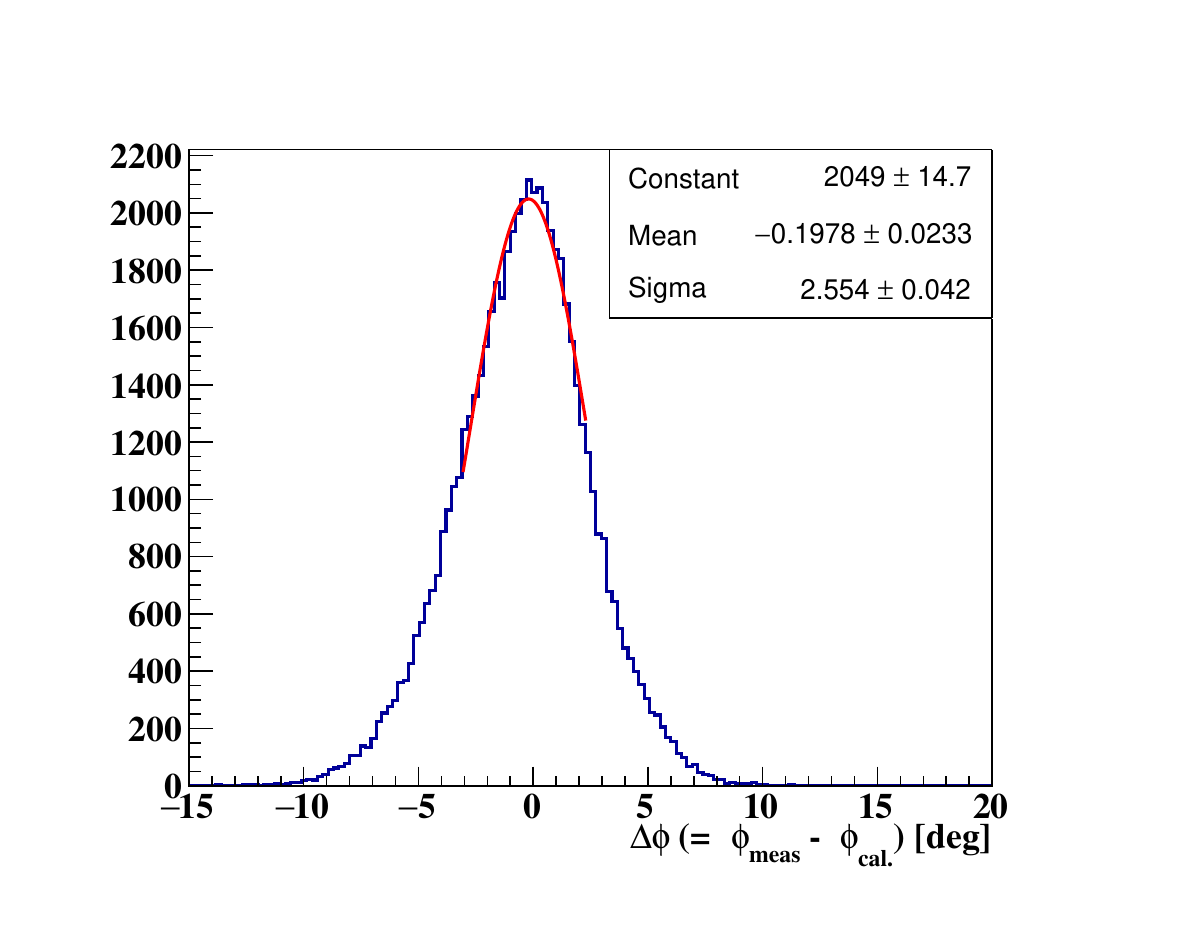}
    \put (20,50) {\Large(b)}
    \end{overpic}

    \begin{overpic}[scale=0.4]{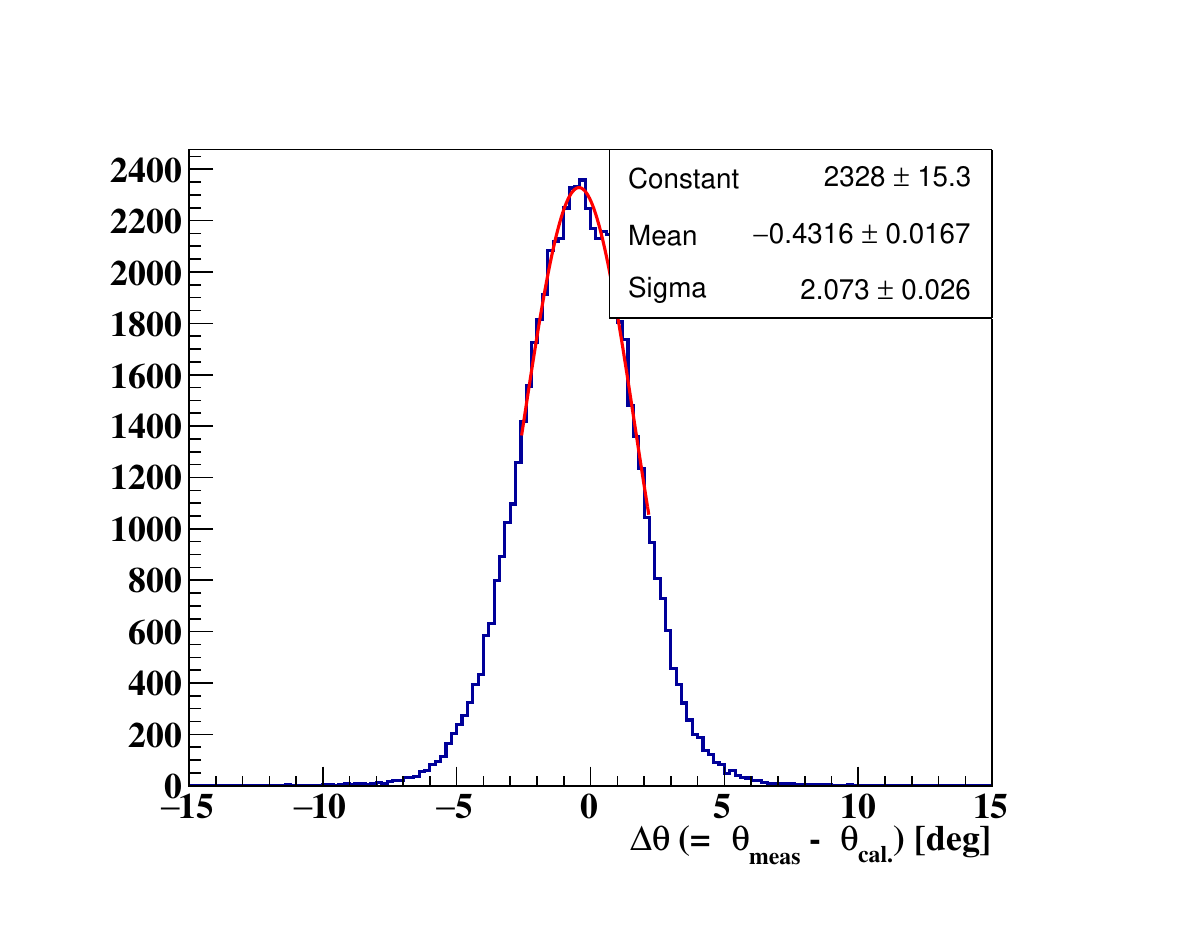}
    \put (20,50) {\Large(c)}
    \end{overpic}
    \begin{overpic}[scale=0.4]{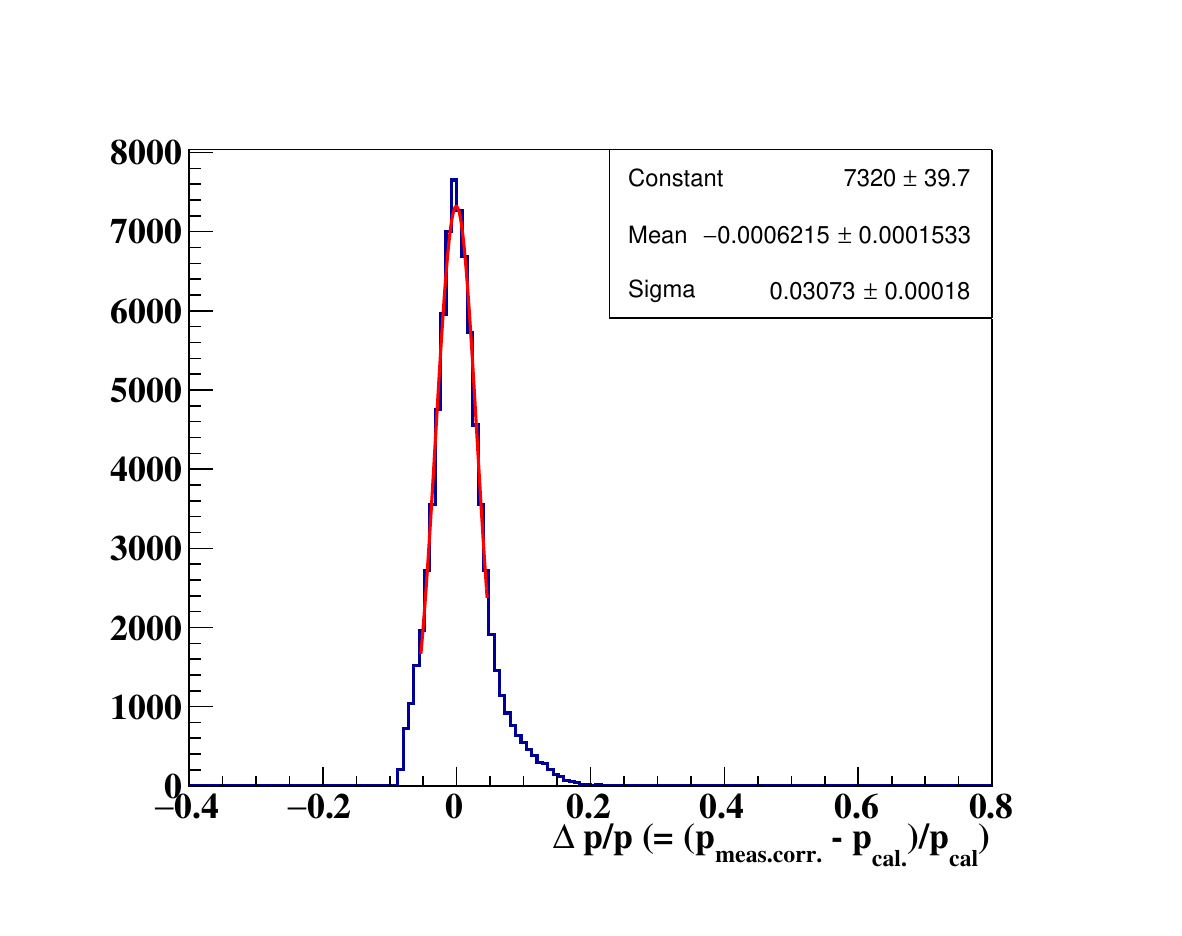}
    \put (20,50) {\Large(d)}
    \end{overpic}
    
    \caption{The measured z-vertex (a), azimuthal angle (b), polar angle(c), and momentum (d) resolutions of the BONuS12 RTPC using elastic $^4$He events at 2.1~GeV electron beam.}
    \label{fig:resolutions_elastic_He4}
    \end{figure*}

In this section, the measured resolutions from the BONuS12 RTPC are measured. Fig.~\ref{fig:resolutions_elastic_He4}(a) shows the measured resolution on the reconstructed z-vertex, Fig.~\ref{fig:resolutions_elastic_He4}(b) shows the azimuthal angle ($\phi$), while Fig.~\ref{fig:resolutions_elastic_He4}(c) shows the measured resolution on the reconstructed polar angle ($\theta$), and Fig.~\ref{fig:resolutions_elastic_He4}(d) shows the momentum resolution using elastic $e^4$He events from our 2.1~GeV electron beam data set. In these plots, the measured $^4$He kinematics in the RTPC are compared to the true $^4$He kinematics that are calculated using the detected scattered electron after applying a set of cuts to select elastic events. Hence, the measured resolutions on the shown quantities are the combined CLAS12 and RTPC resolutions. The observed resolutions are: 1.2~cm $z$-vertex resolution, 2.6$^{\circ}$ resolution on the angle $\phi$, 2$^{\circ}$ resolution on the angle $\theta$, and 3$\%$ resolution on the reconstructed momentum. 

\begin{figure}[t!]
    \centering
    \includegraphics[scale=0.4]{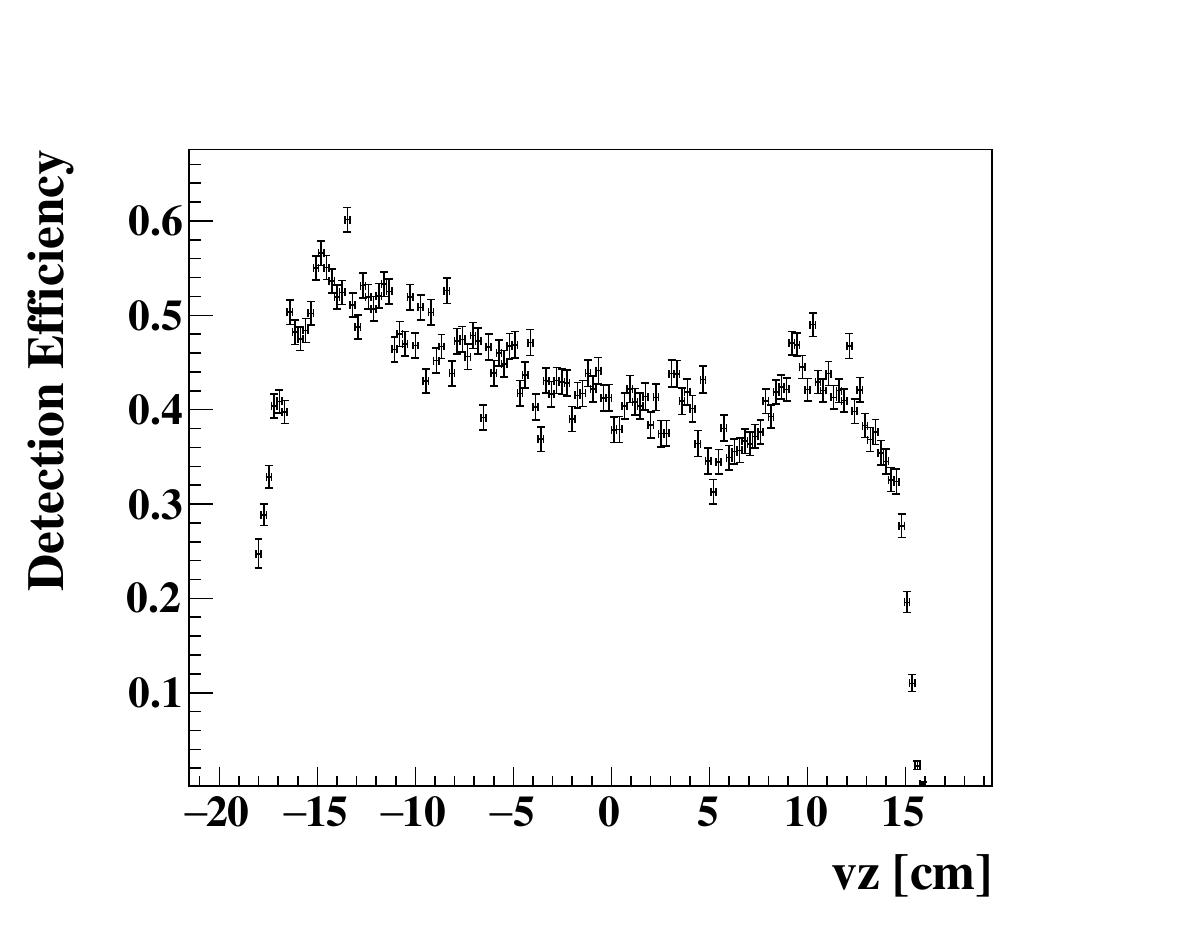}
    \includegraphics[scale=0.4]{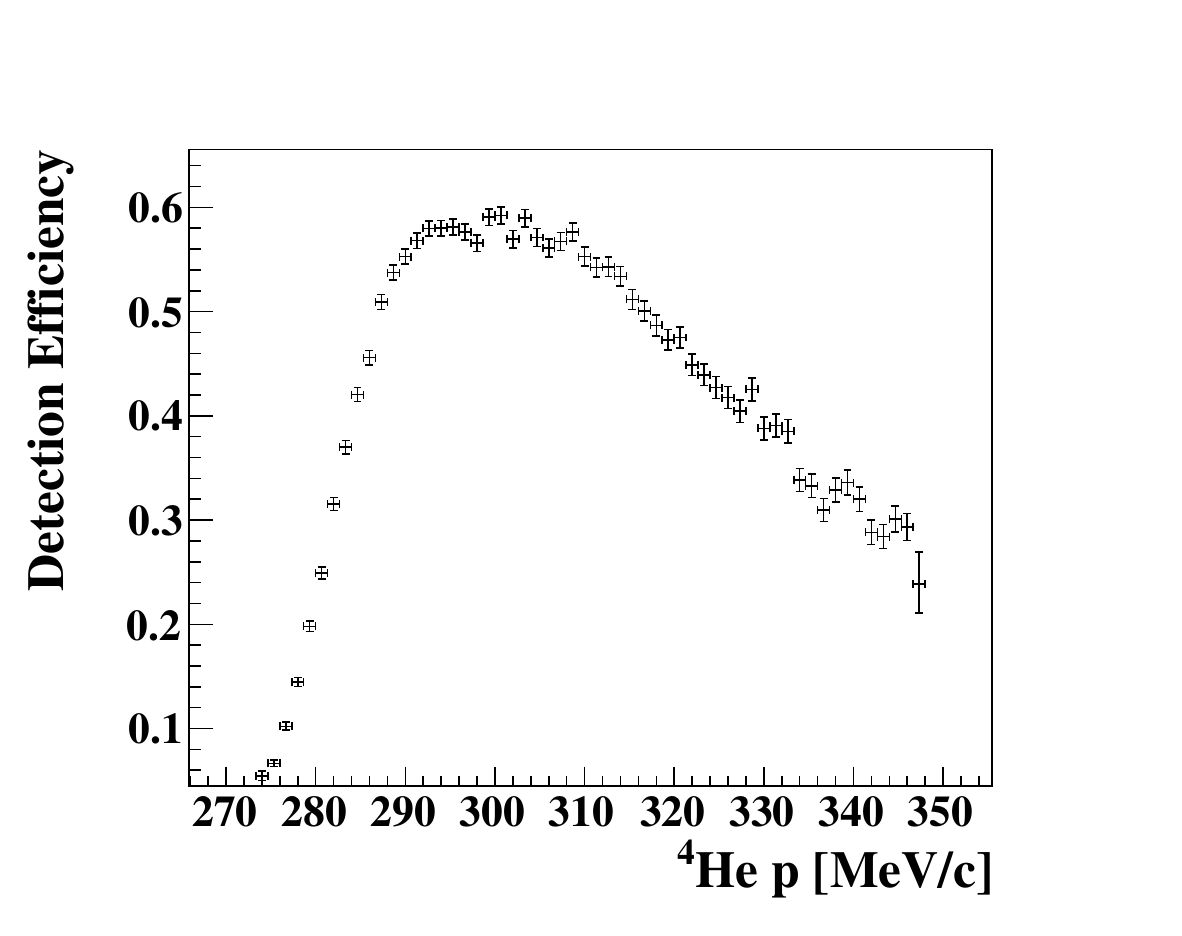}
    \caption{ The measured RTPC detection efficiency as a function of position along the detector (on top) and as a function of recoil momentum (on bottom). These efficiencies are measured using the elastic $^4$He events at 2.1~GeV electron beam. }
    \label{fig:efficiency_elastic_He4}
    \end{figure}

The detection efficiency of the RTPC was studied using the same elastic $e^4$He data set, where the efficiency is defined as the ratio between the number of the detected elastic $e^4$He events to the number of expected elastic event based on the scattered detected electron alone. The RTPC detection efficiency is shown in Fig.~\ref{fig:efficiency_elastic_He4} as a function of the longitudinal position along the RTPC (top) and as a function of the recoiling $^4$He total momentum (bottom). The RTPC detection efficiency is 50$\%$ on average, with about 5$\%$ variation along the detector as a result of the non-uniform charge collection efficiency of the readout pads. The RTPC detection efficiency shows the highest efficiency for the momentum range between 290~MeV/c and 320~MeV/c. At smaller momenta, the recoiling particle would be depositing a large amount of its energy through ionization and would be stopped before making it all the way to the end of the drift region, while in the higher momentum tail, a particle would deposit smaller amounts of energy per ionization hit, in hits that are further apart, making it harder to reconstruct the track efficiently.          

\subsection{\label{sec:operation} Operational Experience
During the Experiment} 
Many lessons have been learned while operating the BONuS12 RTPC during the experiment:

\begin{itemize}
    \item \underline{Gas tightness:} The detector must be gas-tight enough to build up a slight overpressure inside the drift region. In addition, any leaking $^4$He gas from the detector could spread to the neighboring photomultiplier tubes in the other sub-systems and degrade their performance. To alleviate the latter, a set of venting pumps was installed upstream of the detector assembly as an additional precaution to remove the leaking helium and discharge it outside Hall B. 
    
    \item \underline{High-Voltage operation:} the RTPC was operated at two sets of HV values across the GEM foils, 385 V and 375 V, and for a voltage difference between the cathode and the anode (first GEM foil) set to 4300 V. The voltage setting acts as a filter to the types of recoils the detector sees. This is due to the larger signal size for higher GEM voltages, combined with the hardware threshold of the DREAM FEUs. In order to avoid creating big induced currents across the GEM foils while ramping up the HV, the HV was ramped up at a steady pace starting from the cathode and going outwards to the outer side of the third GEM foil. Additionally, induced current limits were implemented on the measured currents at the different layers of the HV system to protect the detector from over-currents in case of beam excursions. Overall, the RTPC proved to be quite stable, with at most a few high voltage trips per 8-hours shift.

    \item \underline{Magnetic field setting:} The optimal operating magnetic field setting was measured empirically using a 10~nA electron beam on a 5.8 atm pressurized H$_2$ target. A magnetic field scan was performed using uniform magnetic field settings ranging from 3 to 5~Tesla. The final value of the magnetic field at the center of the solenoid was chosen based on the detector performance in terms of several factors, i.e., the average number of hits per event detected by the readout pads, the average number of readout pads with collected charge per event, the time distribution of the hits, and the track reconstruction efficiency for the detector using the same electron beam, HV setting, and target gas at different magnetic fields. The electromagnetic background in the forward drift chambers of CLAS12 had to be at a reasonably low level as well. Our studies showed that the best performance of our detector was at 3.78~Tesla of the solenoid field. The dependence on the magnetic field is constrained from above by the larger Lorentz angle of the drifting ionization electrons in the RTPC, and from below by higher background rates in the forward CLAS12 detector.   
    
    \item \underline{Beam centering:} The target straw was 6~mm in diameter and about 50~cm long, making it the longest target used in CLAS12 experiments so far. Beam centering during the BONuS12 experiment was performed in three steps. First, a beam of a few nA current was sent to a dump upstream of the target to check the beam width. Second, a beam of the same current was sent through the target cell. Then, the beam width was measured at locations before and after the target to make sure that the beam was confined within the target cell. The width limits for an acceptable beam are in the range from 100~$\mu$m to 200~$\mu$m for one standard deviation in the $x$ and $y$ profiles of the beam. Finally, the beam was centered at the target straw cell using a Beam Offset Monitor (BOM) system that consists of a set of fiber optics installed at the upstream side of the target. The BOM can be seen on the left side of the target in the drawing shown in Fig.~\ref{fig:target_cad}. The beam was centered by adjusting the beam positions horizontally and vertically and choosing a position with symmetric BOM rates in the full azimuth.  

   \begin{figure*}[ht!]
    \centering
    \includegraphics[scale=1.7]{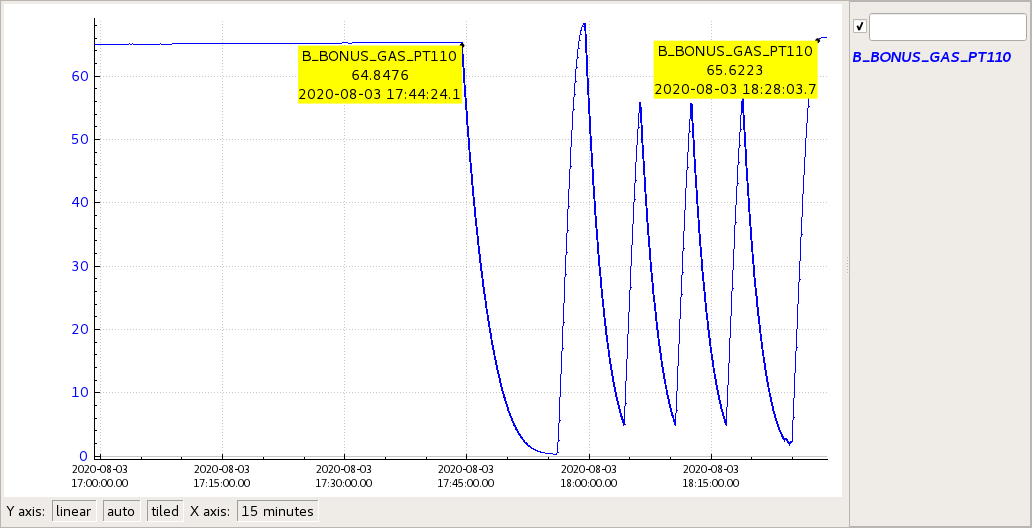}
    \caption{The timeline for the BONuS12 target's pressure during the change of the target gas from D$_2$ to H$_2$. }
    \label{fig:gas_change_D2_H2}
    \end{figure*}

    \item \underline{Noise reduction:} To optimize the signal-to-noise ratio for each RTPC channel, several methods were used. First, the HV for all the GEMs and the cathode were optimized to reliably detect protons while suppressing single-hit noise. Secondly, the common mode noise rejection option of the DREAM chip was selected, in which case only pedestal-subtracted ADC samples above a specific threshold are accepted. Finally, in the reconstruction, chains of hits with fewer than six hits were removed to make sure that the reconstructed tracks had enough ionization hits to be
    cleanly identified. As a result, most events contained only clean tracks and no background hits, as shown in Fig.~\ref{fig:exp_tracks}.

    \item \underline{Target gas purity and purging:} As discussed previously, the BONuS12 experiment used gaseous D$_2$ as the production target, but frequently took data on H$_2$, $^4$He and empty targets for the purpose of background and normalization studies. Therefore, the target gas was changed frequently while maintaining the purity of each gas type. The two main concerns were the outward leakage of the target gas into the detector and the inward helium leakage from the detector into the target gas in case of H$_2$, D$_2$ and empty targets. To address the first issue, the outward leakage was monitored by closing the feeding and the venting lines and measuring the pressure inside the target cell. An average leak rate of 5\% per hour at 68 psig pressure inside the gas targets was frequently measured during the experiment. Regarding the inward leakage from the detector to the target, the purity of the target gas was ensured by purging the target gas at least twice a day. Finally, while changing target gases from one type to another, the target  was filled and purged with the new target gas for a few cycles before resuming data taking. For instance, Fig.~\ref{fig:gas_change_D2_H2} shows the case of the target filled with D$_2$ gas at 65~psig. The gas was vented to zero psig and then filled with 65~psig H$_2$. At this point, the D$_2$ content is less than 10$\%$. This cycle of filling and purging was repeated three times before finally filling with the desired target gas. With this pattern, the D$_2$ contamination in the H$_2$ target gas would be less than 3$\%$, which is the confirmed contamination ratio observed from the data analysis.   

\end{itemize}

\section*{Conclusion}
This paper reported on the construction, operation, and calibration of a next-generation RTPC designed to measure low-energy small-mass recoil particles in a high-rate environment. The operation of the detector was successful and allowed to detect recoil particles with about 50\% detection efficiency at a readout rate of 2~kHz triggered by the detection of high-energy electrons with Jefferson Lab's CLAS12 spectrometer in Hall B. The data collected in 2020 are presently under analysis.

\section*{Acknowledgments}
The authors wish to acknowledge the support of the Jefferson Lab Hall B, Accelerator, and Fast Electronics staff. The authors are particularly indebted to the Hall B engineer, R. Miller, and designers, M. Zarecky and C. Wiggins, as well as C. Cuevas and M. Taylor from the fast electronics group and D. Insley. The authors also wish to thank H. Fenker and S. Stepanyan for useful discussions on the design and construction techniques of previous generation detectors. Finally, the authors are very grateful for the help (both material and advice) given to us from the Micromegas group at Saclay. This material is based upon work supported by the Southeastern Universities Research Association operates the Thomas Jefferson National Accelerator Facility for the United States Department of Energy under contract DE-AC05-06OR23177, and the U.S. Department of Energy, Office of Science, Office of Nuclear Physics, under contract number DE-AC02-06CH11357. This work has received funding from the European Research Council (ERC) under the European Union’s Horizon 2020 research and innovation program (Grant agreement No. 804480).




\bibliographystyle{unsrt}
\bibliography{bonus12rtpc.bib}

\end{document}